\documentclass[english,aps,prx,twocolumn,superscriptaddress,longbibliography]{revtex4-1}
\usepackage[T1]{fontenc}
\usepackage[latin9]{inputenc}
\usepackage{tabularx}
\usepackage{graphicx,wrapfig}
\usepackage{babel}
\usepackage{layouts}
\usepackage{graphicx}
\usepackage{epstopdf}
\usepackage{color}
\usepackage{boxhandler}
\usepackage{subfigure}
\usepackage{dcolumn}
\usepackage{bm}
\usepackage{amsmath}
\usepackage{comment}
\usepackage{bbold}
\DeclareMathOperator{\di}{d\!}

\usepackage{color}

\makeatletter
\@ifundefined{textcolor}{}
{%
 \definecolor{BLACK}{gray}{0}
 \definecolor{WHITE}{gray}{1}
 \definecolor{RED}{rgb}{1,0,0}
 \definecolor{GREEN}{rgb}{0,1,0}
 \definecolor{BLUE}{rgb}{0,0,1}
 \definecolor{CYAN}{cmyk}{1,0,0,0}
 \definecolor{MAGENTA}{cmyk}{0,1,0,0}
 \definecolor{YELLOW}{cmyk}{0,0,1,0}
 }

\newcommand{\bit}{\begin{itemize}}
\newcommand{\eit}{\end{itemize}}

\newcommand{\bea}{\begin{eqnarray}}
\newcommand{\eea}{\end{eqnarray}}

\newcommand{\be}{\begin{equation}}
\newcommand{\ee}{\end{equation}}

\newcommand{\ket}[1]{\left|{#1}\right\rangle}
\newcommand{\bra}[1]{\left\langle{#1}\right|}

\makeatother

\begin{document}


\title{Superconductivity and other phase transitions in a hybrid Bose-Fermi mixture formed by a polariton condensate and an electron system in two dimensions}


\author{Ovidiu Cotle\c{t}}
 \email{ocotlet@phys.ethz.ch }
 \affiliation{Institute of Quantum Electronics, ETH Z{\"u}rich, CH-8093, Z\"urich, Switzerland }
 \author{Sina Zeytino\v{g}lu}
 \affiliation{Institute of Quantum Electronics, ETH Z{\"u}rich, CH-8093, Z\"urich, Switzerland }
\affiliation{Institute for Theoretical Physics, ETH Z{\"u}rich, CH-8093, Z\"urich, Switzerland }

\author{Manfred Sigrist}
\affiliation{Institute for Theoretical Physics, ETH Z{\"u}rich, CH-8093, Z\"urich, Switzerland }
\author{Eugene Demler}
\affiliation{Physics Department, Harvard University, Cambridge, Massachusetts 02138, USA }

\author{Ata\c{c} Imamo\v{g}lu}
\affiliation{Institute of Quantum Electronics, ETH Z{\"u}rich, CH-8093, Z\"urich, Switzerland }

\begin{abstract}
Interacting Bose-Fermi systems play a central role in condensed
matter physics. Here, we analyze a novel Bose-Fermi mixture formed
by a cavity exciton-polariton condensate interacting with a
two-dimensional electron system. We show that that previous
predictions of superconductivity~[F.P. Laussy, Phys. Rev. Lett. {\bf 10}, 104 (2010)] and
excitonic supersolid formation~[I.A. Shelykh, Phys. Rev. Lett. {\bf 14}, 105 (2010)] in this
system are  closely intertwined-- resembling the predictions for  strongly correlated electron systems such as high temperature superconductors. In stark contrast to a large majority of Bose-Fermi
systems analyzed in solids and ultracold atomic gases, the
renormalized interaction between the polaritons and electrons in our
system is long-ranged and strongly peaked at a tunable
 wavevector, which can be rendered incommensurate with the Fermi
momentum. We analyze the prospects for experimental observation of
superconductivity and find that critical temperatures on the order
of a few Kelvins can be achieved in heterostructures consisting of
transition metal dichalcogenide monolayers that are embedded in an
open cavity structure. All  optical control of
superconductivity in semiconductor heterostructures could enable the
realization of new device concepts compatible with semiconductor
nanotechnology. In addition the possibility to interface quantum
Hall physics, superconductivity and nonequilibrium polariton
condensates is likely to provide fertile ground for investigation of
completely new physical phenomena.
\end{abstract}


\maketitle

\section{Introduction}
Interacting Bose-Fermi systems are regarded as a promising platform
for investigating novel many-body physics. Recent advances
demonstrating mixtures of ultracold bosonic and fermionic atomic
gases have intensified research efforts in this class of systems.
Feshbach resonances in two-body atomic collisions can be used to tune the strength of interactions
into the strong coupling regime, allowing for the investigation of
competition between various phase transitions such as supersolid
formation and superconductivity. Motivated by two recent proposals,
we analyze a solid-state Bose-Fermi mixture formed by an
exciton-polariton Bose-Einstein condensate (BEC) interacting with a two dimensional
electron system (2DES). Unlike most solid-state systems, the
interaction strength between the polaritons and electrons can be
controlled by adjusting the intensity of the laser that drives the
polariton system. We find that this system can be used to reach the
strong-coupling regime, evidenced by dramatic softening of the
polariton dispersion at a tunable  wavevector.

Before proceeding, we remark that the interactions between a 2DES
and an indirect-exciton BEC has been
theoretically shown to lead to the formation of an excitonic
supersolid~\cite{shelykh2010rotons,matuszewski2012exciton}. This prior work however, did not take into account the effect of the exciton BEC on the
2DES. Concurrently, the effect of a polariton condensate on a 2DES
has been investigated without considering the back-action of
electrons on the polaritons and it has been predicted that the 2DES
can undergo a superconducting phase
transition~\cite{laussy2010exciton,laussy2012superconductivity,cherotchenko2014superconductivity}.
Our work unifies and extends the above mentioned prior work and
shows that the predicted phase transitions are closely intertwined.

The central finding of our work is  that when screening effects are
taken into account, the long-range polariton-electron interaction is
peaked at a wavevector $q_0$ that is determined by the distance
between the 2DES and the quantum well (QW) hosting the polaritons.
Remarkably, increasing the polariton condensate occupancy by
increasing the resonant laser intensity leads to a substantial
softening of the polariton dispersion at $q_r$ (near $q_0$) which in turn
enhances the strength of the polariton-electron interaction, making it even more strongly peaked. Leaving
a detailed analysis of competition between superconductivity and
potential charge density wave (CDW) state associated with polariton
mode softening as an open problem, we focus primarily on the  superconducting
phase transition.

After introducing the system composed of a bosonic polariton
condensate interacting with a 2DES in Section
\ref{sec:InitialHamiltonian}, we investigate its strong coupling
limit in Section \ref{sec:ManyBody}. Here, we summarize the effects
of many body interactions, leaving the more detailed calculations to
the Appendix \ref{app:ManyBody}. In Section \ref{sec:Results} we use
the theoretical framework developed earlier and analyze the
interactions between a polariton condensate and a 2DES
self-consistently. We notice that the strong interactions can lead
to instabilities both in the condensate and in the 2DES. We
investigate quantitatively the instability of the 2DES towards
superconductivity~
\cite{laussy2010exciton,laussy2012superconductivity,cherotchenko2014superconductivity}
while also taking into account the effect of the 2DES on the BEC. We
also comment briefly on the instability of the 2DES towards the
formation of an unconventional CDW ordered state as a consequence of
the renormalized electron-polariton interaction becoming strongly
peaked at wavevector $q_r$. In Section \ref{sec:Materials} we
investigate how to reach the strong coupling regime experimentally
in order to observe these phase transitions. We find that
superconductors with temperatures of a few Kelvins can be obtained
in transition metal dichalcogenide (TMD) monolayers. We briefly
summarize our results and provide { a short description of new physics and applications enabled by our analysis in Section
\ref{sec:Conclusion}.}


\section{Theoretical investigation}

\subsection{Description of the coupled electron-polariton system}\label{sec:InitialHamiltonian}
\begin{figure}[t!]
  
   \includegraphics[width=0.9\columnwidth]{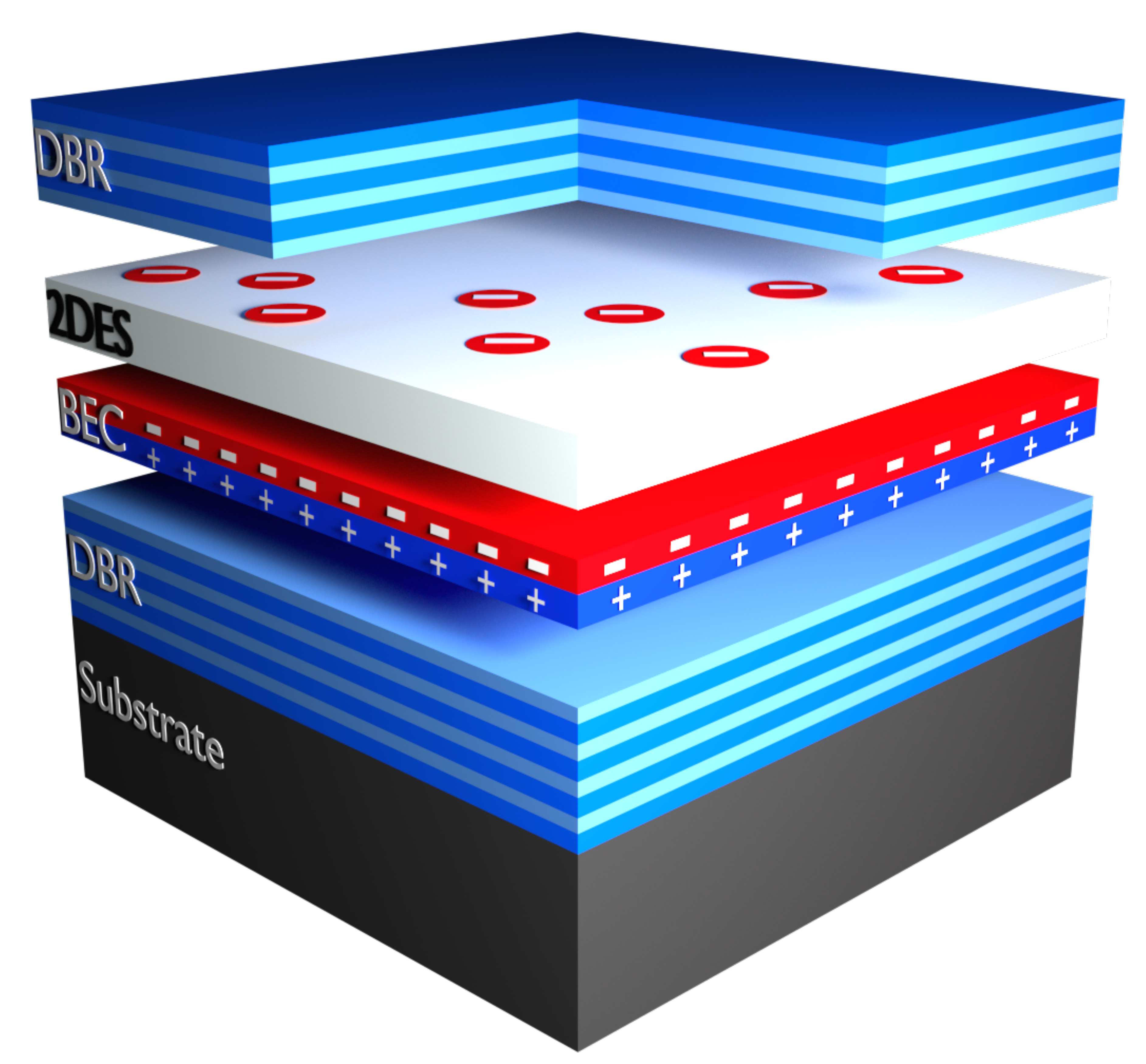}
    \includegraphics[width=\columnwidth]{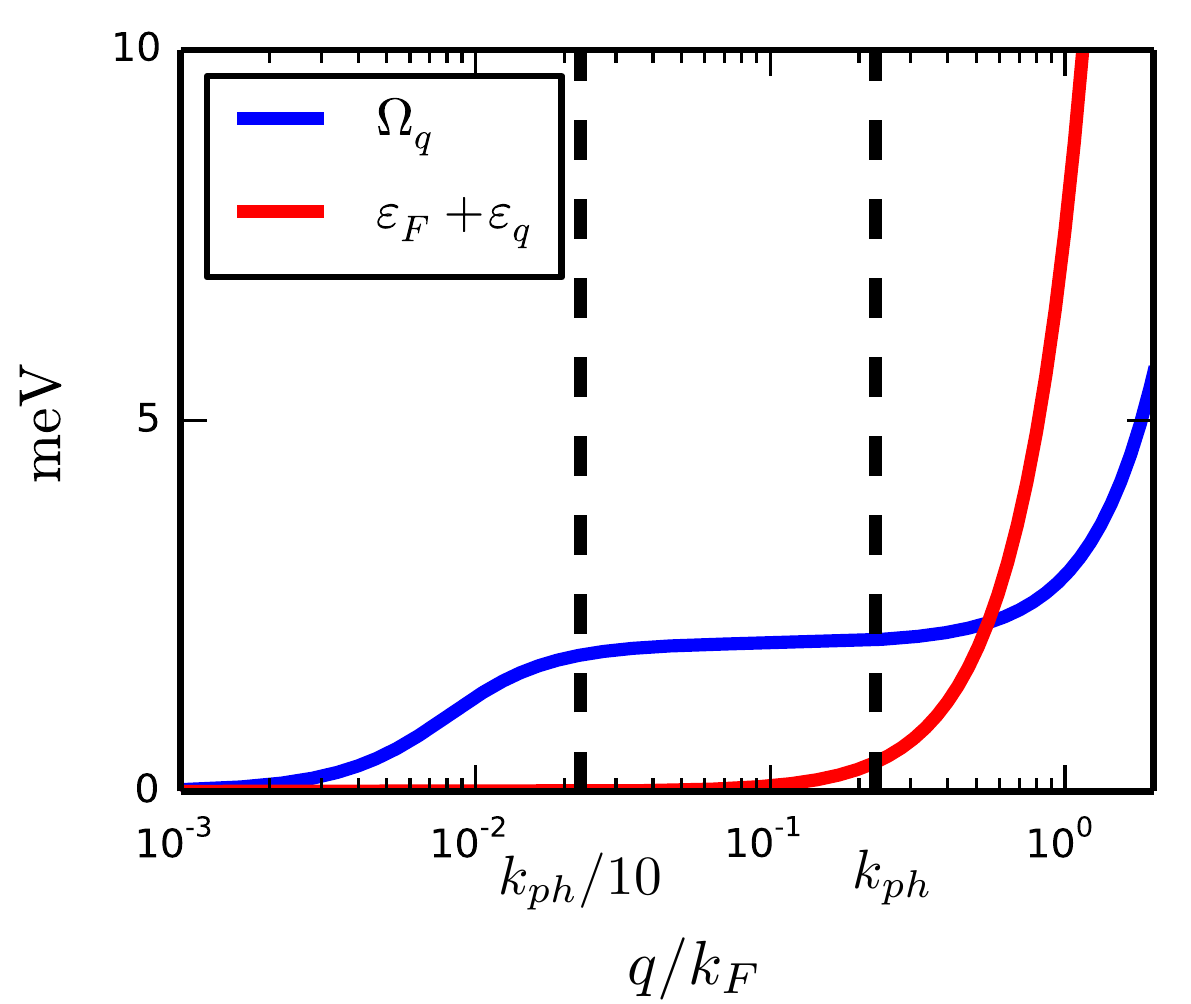}

  \caption[]{Upper panel:  The schematic of the semiconductor heterostructure that is analyzed.  Lower panel: bare polariton (blue) and electron (red) dispersion in logarithmic scale. The vertical dashed lines are at the photon wavevector $k_{ph}= 3.3E_c(0)/(\hbar c)$  (corresponding to the maximal momenta that we can investigate optically; the $3.3$ factor comes from the GaAs index of refraction) and at $k_{ph}/10$ (roughly corresponding to the momentum where the polariton dispersion switches from photonic to excitonic). The parameters used are typical GaAs parameters: $g_0=2\mathrm{meV}$,  $m_e=0.063m_0$, $m_h=0.046m_0$, $n_e = 2\times10^{11}\mathrm{cm}^{-2}$, $E_c(0)=E_\mathrm{x}(0)=1.518\mathrm{eV}$. }
   \label{fig:Setup}
\end{figure}

The system that we investigate is similar  to the system in
Ginzburg's proposal for high temperature exciton-mediated
superconductors~\cite{ginzburg1964surface}. It consists of a 2DES in
close proximity to a quantum well (QW) in which excitons can be
created by shining a laser resonantly without influencing the 2DES. The whole system  is embedded in a
cavity formed by a pair of distributed Bragg reflectors (DBRs),
which confine light and { and allow for a strong
interaction between excitons and photons. Due to the
non-perturbative light-matter coupling the new eigenstates are
composite particles called polaritons}. The polaritons can form a
BEC either under non-resonant or under direct resonant excitation by
a laser~\cite{carusotto2013quantum}. The interaction between neutral polaritons and the electrons in the 2DES is due to the excitonic content of the polaritons and can be enhanced by enhancing the size of the dipole of the exciton using a DC electric field. In this
scenario, an attractive interaction between electrons can be
mediated by the polariton excitations of the BEC. As we will show
below, the strength of the interaction is proportional to the number
of polaritons in the condensate which can be tuned experimentally.
The schematic design of the experimental setup is presented in the upper
panel of Figure~\ref{fig:Setup}.

In the following we will assume for
simplicity that the polariton condensate is generated by a resonant
laser field for the $k=0$ state of the lower polariton branch. The Hamiltonian describing the coupled exciton and
cavity-mode dynamics can be diagonalized through a canonical
transformation and the resulting lower and upper polariton
eigenstates are superpositions of exciton and photon states. Since the
whole system is translationally invariant, in-plane momentum $k$ is
a good quantum number for polaritons. The annihilation operator for
a lower polariton of momentum $k$ is: 
\bea 
b_k = X(k) a^{(\mathrm{x})}_k + \sqrt{1 - X(k)^2} a^{(c)}_k, 
\eea
where $a^{(\mathrm{x})}$ and $a^{(c)}$ are the exciton and cavity-photon
annihilation operators. $X(k)$ is the exciton fraction of the lower
polariton mode with  wavevector $k$ and is given by: 
\bea 
\left|X(k) \right| ^2 =\frac{1}{2} \left( 1 - \frac{\delta E(k)}{\sqrt{\delta E^2(k)+4 g_0^2}} \right), 
\eea 
where $g_0$ is the light-matter coupling strength and $\delta E(k)=E_{\mathrm{x}}(k) -
E_c (k) $ is the energy difference between the exciton and the
cavity mode.  Denoting the $k=0$ detuning between the photon and exciton
dispersion by $\Delta$ we express the energy difference: $\delta
E(k) =\Delta+ \hbar^2 k^2 \cdot ( m_\mathrm{x}^{-1} - m_c^{-1}) /2 $ where
$m_e$, $m_h$, $m_\mathrm{x}=m_e+m_h$, $m_c$ denote the electron, hole,
exciton, and cavity effective masses. We also define the polariton
mass $m_p$ by $m_p^{-1}= m_\mathrm{x}^{-1}+m_c^{-1}$.

Since the upper branch polaritons are unstable against relaxation into
lower energy polariton states, we focus exclusively on excitations
within the lower polariton branch within which a BEC
in the k=0 mode is formed. First we
write the initial Hamiltonian of our system as
\bea \label{GeneralH}
H &=& H_{0}^{(e)}+H_{0}^{(p)}+H_{I}^{(e-e)}+H_{I}^{(e-p)}+H_{I}^{(p-p)},\nonumber\\
H_0^{(e)} &=& \sum_k \varepsilon_k c_k^\dagger c_k,\quad H_0^{(p)} =\sum_k \Omega_k b_k^\dagger b_k,\nonumber\\
H_I^{(e-e)}&=&\frac{1}{2} \sum_{k,k',q} V_C(q) c_k^\dagger c_{k'}^\dagger c_{k+q} c_{k'-q}, \\
H_I^{(p-p)} &=&\frac{1}{2} \sum_{k,k',q} U(k,k',q) b_k^\dagger b_{k'}^\dagger b_{k+q} b_{k'-q}, \nonumber \\
H_I^{(e-p)} & = & \sum_{k,k',q} V_X(k,q) b_k^\dagger c_{k'}^\dagger b_{k+q} c_{k'-q},  \nonumber 
\eea
where $c^\dagger_k$ ($c_k$) denote
the electron creation (destruction) operators and $V_C(q)=e^2/(2 A
\epsilon q)$ is the usual Fourier transform of the Coulomb
interaction between electrons. ($A$ is the normalization area, $e$
is the electron charge and $\epsilon$ is the dielectric constant of
the medium.)

The electron and (lower) polariton dispersions are given by:
\bea
\varepsilon_k &=& \frac{\hbar^2 k^2}{2 m_e} - \varepsilon_F,\\
 \Omega_k &=&\frac{1}{2}\left(\sqrt{\Delta^2+4 g_0^2} + \frac{\hbar^2 k^2}{2 m_p} -  \sqrt{\delta E^2(k) +4 g_0^2}\right). \nonumber
 \eea
 where $\varepsilon_F$ is the Fermi energy of the 2DES. Given the above parabolic dispersion, $\varepsilon_F$ is proportional to the electron density $n_e$. Unless otherwise stated, for the rest of the paper we will set $\Delta=0$. For this case, the polariton dispersion is plotted in logarithmic scale in blue in the lower panel of Figure \ref{fig:Setup}. For comparison, we also superimpose the electron dispersion in red in the same panel. 

The polariton-polariton interaction is given by
\bea
U(k,k',q) = X(k) X(k')X(k+q) X(k'-q) \frac{U}{A}. 
\eea 
As expected, this  interaction is
proportional to the exciton fraction $X(k)$ of the polaritons
involved. We assume a simple contact interaction between excitons
even though this interaction is more complicated than the above equation suggests~\cite{ciuti1998role,byrnes2010mott,byrnes2014effective}. We also emphasize that
polariton-polariton interaction depends on the polarization of the
polaritons and is potentially tunable through the use of Feshbach
resonances~\cite{carusotto2010feshbach,takemura2014polaritonic}. In
our analysis we will treat $U$ as a freely tunable parameter.

As mentioned above, excitons are neutral particles and therefore do not couple  strongly
to electrons but a significant electron-polariton interaction
can be created by inducing a dipole in the excitons. This can be
done in various ways depending on the choice of the material. For
example, one can use an electric field perpendicular to the exciton
plane to polarise the excitons, however, this will result in small dipoles. A better approach is to  use two tunnel-coupled QWs and bias the
energy bands in such a way that holes can live only in the first QW,
while electrons can freely tunnel between the QWs. This leads to a new type of
polariton which has both a large dipole (from the indirect exciton
part) and a large light matter coupling (from the direct exciton
part). In this way polaritons with dipoles of lengths comparable to
the exciton Bohr radius, known as dipolaritons, can be
produced~\cite{cristofolini2012coupling}.

Regardless of the mechanism, in the  approximation of infinitely
thin QWs, one can obtain an analytical expression for the
electron-polariton interaction $V_X$ as shown in Ref.
\onlinecite{laussy2012superconductivity}. We will use this analytical expression in numerical simulations but below we show how to obtain an approximate but simpler expression.   First, $V_X$ is
electrostatic in origin and it will be proportional to $V_C(q)$.
Since the interaction is due to the partially excitonic nature of polaritons it
will be proportional to the exciton fraction of the polaritons
involved. Finally, $V_X$ is proportional to the exciton dipole
length $d$. We expect that polaritons will not be able to respond to
momentum transfers larger than $1/a_B$ ($a_B$ denotes the exciton
Bohr radius), $1/d$ and $1/L$,
where $L$ denotes the distance between the 2DES and the position of the centre of mass of the polaritons in the direction orthogonal to the 2D planes.
Analyzing the expression in Ref.
\onlinecite{laussy2012superconductivity} we see that the dominant
momentum cutoff is given by the distance $L$, so in the limit of
$d,a_B \leq L$ it takes the following approximate form
\footnote{We
mention that this expression has been obtained by assuming that the
in-plane exciton wavefunction is not modified by the induced dipole.
A better exciton wavefunction is the one used in Ref.
\onlinecite{leavitt1990excitonic}, unfortunately, this expression is
no longer analytical.}: 
\bea\label{EPH} 
V_X(k,q) \approx X(k) X(k+q) V_C(q) q d e^{-q L} . 
\eea

Since we assume a polariton condensate in the $k=0$
mode,  we follow the Bogolyubov prescription and set $b_0 =
b^\dagger_0 = \sqrt{N_0}$. We denote by $N_0$ ($n_0$) the number (density) of polaritons in the BEC. We then make the Bogolyubov approximation
which consists in ignoring terms of lower order in $N_0$. Leaving
the details to Appendix
\ref{app:ManyBody}, we remark that the polariton-polariton
interaction Hamiltonian in this limit reduces to a quadratic
Hamiltonian with an interaction strength $U(q) = U(0,0,q)$, which
can be eliminated through a canonical
transformation. More importantly, the electron-polariton interaction
Hamiltonian after the Bogolyubov approximation has the same
structure as the electron-phonon Hamiltonian, with a tunable
interaction strength: 
\bea H_I^{(e-p)} = \sum_{k,q} \sqrt{N_0}V_X(q) c_{k+q}^\dagger c_k (b_q + b^\dagger_{-q}),  
\eea 
where $V_X(q) = V_X(0,q)$.

\subsection{Theoretical investigation}\label{sec:ManyBody}
Given the formal correspondence between the electron-phonon and
electron-polariton interaction Hamiltonians, we apply the well-known
Migdal-Eliashberg theory~\cite{mahan2000many,scalapino1969electron}
developed for the electron-phonon Hamiltonian to analyze the
electron-polariton interaction. This theory was developed by
Eliashberg starting from Migdal's theorem, which is the equivalent
of the Born-Oppenheimer approximation in Green's function language.
We find that our system also satisfies Migdal's theorem, which
justifies the use of Migdal-Eliashberg theory.

One of the important results of our theoretical analysis revolve around the substantial softening of the polariton dispersion at a wavevector $q_r$ and the appearance of a roton-like minimum at this wavevector (see upper panel of Figure \ref{fig:RenormalizedPolariton}). We find that the electron-polariton interaction and consequently the effective electron-electron attractive interaction mediated by polaritons increases significantly due the softening of the polaritons and is strongly peaked at the wavevector $q_r$. In the strong-coupling regime, characterised by a significant polariton softening, both the polariton BEC and the 2DES are susceptible to phase transitions.

As far as we know, this strongly peaked interaction in momentum space at a tunable wavevector is unique to our system and stands in stark contrast to the contact interaction in neutral Bose-Fermi mixtures formed with cold-atoms or the Kohn anomaly in solid-state systems that can result in large interactions at twice the Fermi wavevector. We will discuss this strongly peaked interaction in more detail in Section \ref{sec:ResultsGeneral}.
In the following we briefly summarise our results. For a detailed derivation we refer to Appendix \ref{app:ManyBody}.

\subsubsection{Screening due to the electron system}
The reason for the strongly peaked interaction in the momentum space can be traced to the screening by the mobile carriers of the electron system.
The bare electron-polariton interaction has an exponential cutoff
in momentum space due to the distance between the polariton and
electron planes. At the same time, $V_X(q \to 0) \to
\mathrm{constant}$.

When screening is taken into account (the lower panel of Figure \ref{fig:RenormalizedPolariton}) we have to renormalize $ V_X$ as $\tilde{V}_X(q) \to V_X(q) /
\epsilon(q)$ where $\epsilon(q)$ is the static Thomas-Fermi
dielectric function given by $\epsilon(q) = 1 + k_{TF}/q$  where
$k_{TF} = m_e e^2/ 2 \pi \epsilon \hbar^2=2/a_B$. As one can easily
observe, the effect of screening is to cut off the contribution of
small wavevectors such that $V_X(q\to 0) \to 0$. This small momentum
cutoff together with the large momentum cutoff mentioned above leads
to a maximum in the interaction in momentum space at the wavevector
\bea 
q_0= \frac{1}{a_B}  \left[ \sqrt{1+ \frac{2 a_B}{L}}-1\right],  
\eea
as shown in the lower panel of Figure \ref{fig:RenormalizedPolariton} in blue solid line. The broad maximum in the interaction becomes very strongly peaked as the polaritons soften and the electron-polariton interaction approaches the strong-coupling regime.

The position of $q_0$ depends only on the screening wavevector
$k_{TF}$ and consequently on the electron Bohr radius $a_B$, as well as the distance between the BEC and the 2DES. As a consequence, $q_0$
can be tuned by changing the distance between the BEC and the
2DES. We show the tunability of this interaction by changing
the distance $L$ in the lower panel of Figure
\ref{fig:RenormalizedPolariton} in dashed blue line. Alternatively, one can tune the
relative position of $q_0$ with respect to the Fermi wavevector
$k_F$ by changing the 2DES Fermi energy. This latter method can
provide real-time control of $q_0/k_F$.

\subsubsection{Polariton softening}
Remarkably, as the electron-polariton interaction increases  the
polaritons tend to soften due to interactions with the electrons.
Since the interaction is already weakly peaked in momentum space this softening will be most
drastic around a certain wavevector  $q_r$ which we refer to as a roton-like minimum. As we show below, this softening results in a significant increase in the electron-polariton
interaction without which the strong-coupling regime cannot be
reached.

The effect of electron-polariton interactions on the polariton
spectrum can be  understood as a renormalization of the interaction
between a polariton at momentum $q$ and a polariton in the
condensate. In linear response the correction to the polariton-polariton interaction is given by $\chi(q)
V_X^2 (q)$ where $\chi(q)=\chi_0(q)/\epsilon(q)$ is the response
function in the random phase approximation (RPA) and $\chi_0(q)$ is
known as the Lindhard function and denotes the linear response of
the electrons in the absence of electron-electron interactions. Since $\chi(q)$ is typically negative, one can imagine the density fluctuations in the 2DES mediating attractive interactions between the polaritons in the BEC.
Intuitively, a polariton of momentum $q$ creates a potential
$V_X(q)$ in the 2DES which responds by creating a charge
distribution $\delta n (q) = \chi(q) V_X(q)$. This in turn attracts
a polariton in the condensate with the strength $\delta n(q) V_X(q)$.

The interacting Bose-condensate can be  exactly diagonalized in the
Bogolyubov approximation which yields the renormalized polariton
dispersion: 
\bea
\label{omegaq} \omega_q \to \sqrt{\Omega_q^2+  2N_0\Omega_q \left[U(q) + \chi (q) V_X^2(q) \right] } .  
\eea

\begin{figure}[t!]
  \centering
   \includegraphics[]{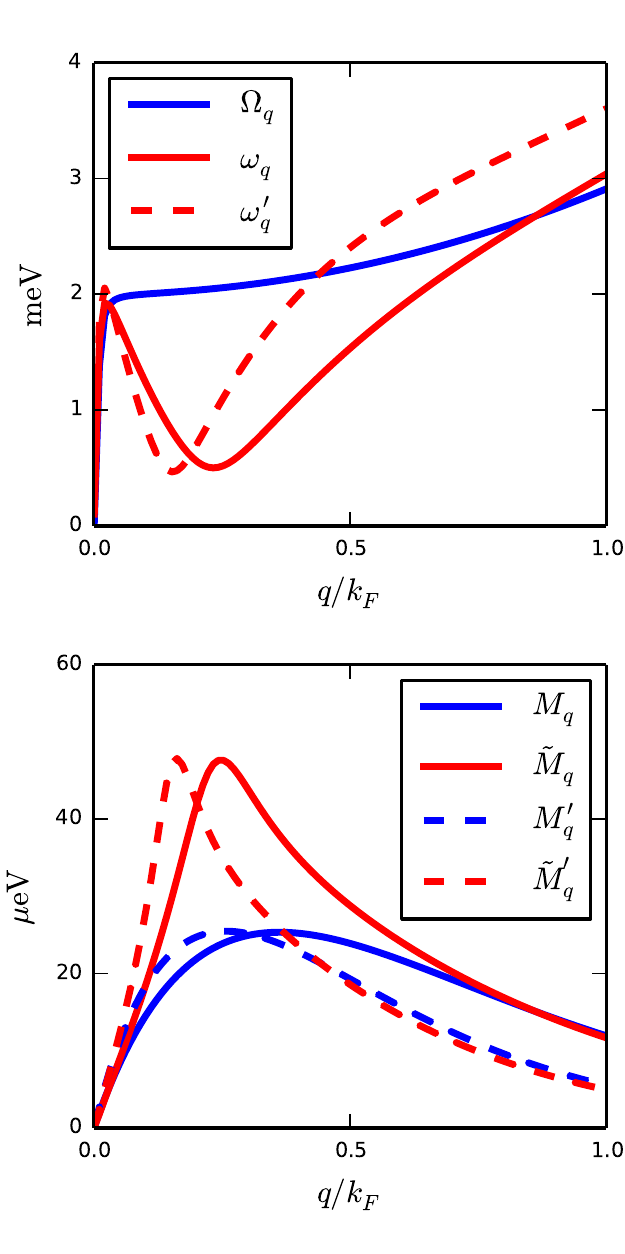}
  \vspace{ -0.2in}
  \caption{Upper panel: bare polariton dispersion (blue) versus renormalised polariton dispersion (red). Lower panel: screened electron-polariton matrix element $M(q)=\sqrt{N_0} V_X(k) / \epsilon(k)$ (blue) versus screened and renormalised electron-polariton matrix element $\tilde{M}(q)$ (red). The parameters used for the solid lines are typical GaAs parameters: $d=a_B=10\mathrm{nm}$, $L=20\mathrm{nm}$, $g_0=2\mathrm{meV}$, $\epsilon=13 \epsilon_0$,  $m_e=0.063m_0$, $m_h=0.046m_0$, $n_e = 2\times10^{11}\mathrm{cm}^{-2}$, $U=0.209\mu\mathrm{eV} \mu \mathrm{m}^2$, $n_0=4 \times 10^{11}\mathrm{cm}^{-2}$. The parameters used for the dashed lines are the same as for the solid lines except for $L'=1.5L$ and $n_0'=2 n_0$.}
   \label{fig:RenormalizedPolariton}
\end{figure}

We plot the renormalized polariton dispersion in the upper panel of
Figure~\ref{fig:RenormalizedPolariton} for some typical GaAs
parameters. Since the $q$ dependences of $U(q)$ and $\Omega_q$ are negligible in comparison to that of $\chi(q)V_X^2(q)$, by maximizing the latter we determine the roton minimum:
\bea 
q_r\approx \frac{1}{a_B}  \left[ \sqrt{1+ \frac{a_B}{L}}-1\right]  
\eea
Note the slight difference between $q_r$ and $q_0$ which stems from the
fact that the exponential cutoff is twice as effective for a second
order interaction.  This softening has been investigated theoretically
for excitons in the context of supersolidity~\cite{shelykh2010rotons,matuszewski2012exciton}.

The electron-polariton interaction also gets renormalised as the polariton dispersion softens. The renormalised electron-polariton interaction matrix element is: 
\bea
\tilde{M}(q) &\to& \sqrt{N_0} \frac{V_X(q)}{\epsilon(q)}\sqrt{\frac{\Omega_q}{\omega_q}}.  
\eea 
We present this renormalized interaction in the lower panel of Figure
\ref{fig:RenormalizedPolariton}. Although the factor of
$\sqrt{\Omega_q/\omega_q}$ which leads to an increase in
interactions may look similar to the renormalization factor encountered in the electron-phonon interaction~\cite{mahan2000many}, it has a slightly different physical origin. In the phonon case, this renormalization factor is due to the quantisation of the position operator of the harmonic oscillator. In the polariton-electron system this factor appears because of the condensate depletion due to interactions as shown in Appendix \ref{app:ManyBody}. Therefore, we can increase interactions by increasing the condensate depletion. However, one must always make sure that the condensate depletion remains small compared to the number of polaritons in the condensate to ensure that Bogolyubov's approximation remain valid. The peak of the renormalized matrix element will be between $q_0$ and $q_r$ depending on the strength of this renormalisation factor.

\subsubsection{Renormalized Hamiltonian}
We can gather all the results from the previous analysis and write down a renormalised Hamiltonian for the new quasi-particles:
\bea
H &=& H_{0}^{(e)}+H_{0}^{(p)}+H_{I}^{(e-e)}+H_{I}^{(e-p)},\nonumber\\
H_0^{(e)} &=& \sum_k \tilde{\varepsilon}_k \tilde{c}_k^\dagger \tilde{c}_k,\quad H_0^{(p)} =\sum_k\omega_k \tilde{b}_k^\dagger \tilde{b}_k,\nonumber\\
H_I^{(e-e)}&=&\frac{1}{2} \sum_{k,k',q} \tilde{V}_C(q) \tilde{c}_k^\dagger \tilde{c}_{k'}^\dagger \tilde{c}_{k+q} \tilde{c}_{k'-q}, \nonumber\\
H_I^{(e-p)} & = & \sum_{k,q} \tilde{M}(q) \tilde{c}_{k}^\dagger \tilde{c}_{k-q} \left(\tilde{b}^\dagger_q + \tilde{b}_{-q}\right),  
\eea where we denoted the new  quasi-particles and
quasi-particle interactions with a tilde. In the above
$\tilde{V}_C(q) = V_C(q) / \epsilon(q)$ and $\tilde{\varepsilon}_k =
\hbar ^2 k^2 / (2 m_e^*)$. We remark that the effect of
polaritons on the electron dispersion is to increase the electron
mass from $m_e$ to $m_e^*$ as shown in Eq.~(\ref{effectivemass}).

We caution that in using the  above Hamiltonian in perturbative calculations one should be careful to avoid double counting  since electron-hole bubble diagrams have already been taken into account.

\subsection{Justification of the renormalized Hamiltonian description}
When using the diagrammatic techniques, we make two
important approximations. The first approximation is to ignore the
finite linewidth of the polariton and electron spectral functions, due to many-body
interactions. We investigate the validity of this approximation in
Section~\ref{finite}. The second and arguably the most important
approximation that we make is to choose which diagrams to discard
and which diagrams to sum over. Our choice was motivated by the
Migdal-Eliashberg theory.

In order to investigate the validity of these approximations, we
first  introduce the Eliashberg function and the
electron-polariton coupling (EPC) constant. In analogy to the
Migdal-Eliashberg theory for phonons the electron-polariton
interaction can be characterised by the EPC constant $\lambda$: 
\bea
\lambda = 2 \int_{0}^{\infty} \frac{\di \omega}{\omega} \alpha^2F(\omega),  
\eea where we introduced the commonly used Eliashberg
function $\alpha^2 F(\omega)$, the only quantity we need to know in
order to assess the effect of the polaritons on the electrons~\cite{mahan2000many}. The
Eliashberg function is related to the scattering probability of an
electron on the Fermi surface through a virtual polariton of
frequency $\omega$: 
\bea
\alpha^2 F(\omega) = \sum_{k,k'} \left| \tilde{M}_{k-k'} \right|^2 \delta(\omega- \omega_{k-k'}) \delta (\varepsilon_k) \delta (\varepsilon_{k'}) /N(0), \nonumber\\ \eea where $N(0)$ is the electron
density of states at the Fermi surface.

The EPC constant quantifies the strength of the electron-polariton coupling and the strong coupling regime is characterised by large values of this parameter. Many properties of the interacting electron-polariton system depend on this constant (for example the electron mass gets renormalised such that $m_e^* = m_e( 1+ \lambda)$).

\subsubsection{Quasi particle approximation}\label{finite}
In the strong coupling regime one has to check whether the quasi-particle description remains valid for electrons, i.e. whether the quasi-particle linewidth is much smaller than the quasi-particle energy.

At zero temperature, according to Eq.~(\ref{ELife}), the electron quasi particle linewidth $\Gamma$ at energy $\omega$ above the Fermi surface is given by:
\bea
\Gamma (\omega) =\pi \int_0^\omega \di \omega' \alpha^2 F(\omega').  
\eea

Generically, as long as the polariton energy scale is much larger than the superconducting gap, the electrons forming the Cooper pairs will be good quasi-particles since they will interact only virtually with polaritons. While this is guaranteed to the extent that polaritons themselves are good quasi-particles, in the limit of substantial polariton softening we need to ensure that $\Gamma(k_B T_c) \ll k_B T_c$, where $T_c$ is the critical temperature of the superconductor. 

If we neglect cavity losses, the polariton excitations can decay only by creating electron-hole pairs. Even if the polariton excitations in the strong-coupling regime are not well-defined excitations, our investigation of the transitions of the 2DES are not affected as long as the electrons close to the Fermi surface remain good quasiparticles. The finite polariton excitations' linewidth can be incorporated in our calculations by changing the Eliashberg function. We comment on this in Appendix \ref{app:ManyBody} and show that this in fact does not influence our results and that the electron quasi-particle picture remains valid since $\Gamma(\omega) \propto \omega^2$ for small frequencies.

In addition to the above there are, of course, intrinsic decoherence mechanisms that appear due to interactions that are neglected when writing down the initial Hamiltonian in Eq.~(\ref{GeneralH}). The most important decoherence mechanism is the scattering by impurities in the system. Impurities will lead to a broadening of the polariton dispersion due to localisation effects. This will limit how much the polaritons can soften. Since impurity induced broadening is typically Gaussian, it can be neglected provided that the polariton energy exceeds the corresponding linewidth. Since the electrons are in a high mobility 2DES we do not expect any disorder/locaziation effects to significantly affect them. 

\subsubsection{A Debye energy for polaritons?}\label{Debye}
In normal metals there is a frequency cutoff for phonons which is
much lower than the  Fermi energy. This energy scale separation is
crucial for theoretical investigations, because it allows one to
make an adiabatic approximation in which electrons instantaneously follow
the lattice motion. In our system there is no energy cutoff;
however, not all polaritons interact as strongly with electrons. The
polaritons that couple most strongly to electrons have energies
bounded roughly by  $\omega_D$, in analogy to
the Debye frequency in the case of phonons. We consider this to be
the relevant energy scale of the polaritons.

As mentioned above, the separation of electron and polariton energy
scales  allows one to make a Born-Oppenheimer approximation (known as
Migdal's theorem in diagrammatic language) and obtain a perturbative expansion
in the small parameter $\hbar \omega_D/\varepsilon_F$. In some materials (as in
GaAs) this condition is already satisfied, without including
renormalisation effects, due to the different electron/exciton
masses. However, in other materials, the electron/exciton masses are
comparable (as in TMD monolayers). In these materials
renormalization effects are crucial in creating a small Debye
frequency and allowing the use of Migdal's theorem for a theoretical
investigation of the system.  

We notice that  the polaritons at the roton minimum will interact
most strongly with the electrons, as shown in Figure
\ref{fig:RenormalizedPolariton}. Furthermore, as we will see in the
next section, the effective electron-electron attraction between the
electrons on the FS is inversely proportional to the frequency of
the polaritons mediating the interaction, which is what one would
also expect from a second order perturbation theory. Therefore, we
expect the typical energy scale of the polaritons that dominate the contributions to  attractive electron-electron interactions, to be of the
order of the polariton energy at the roton minimum.
Quantitatively, we define the following Debye frequency: \bea
\omega_D = 2 \int_{0}^{\infty} \di \omega\ \alpha^2 F(\omega)/
\lambda.   \eea

Notice that for weak coupling the Debye frequency $\omega_D$ will be of the order of the light matter coupling $g_0 / \hbar$. 

\section{Superconductivity and Charge Density Waves}\label{sec:Results}
In the previous section we developed the theoretical framework needed to understand a system of fermions interacting with a bosonic condensate. The theory is general within the framework of the Migdal's theorem and up to the assumption that density-density interactions depend only on the momentum transfer.

In this section,  we restrict our analysis to the system introduced in Section \ref{sec:InitialHamiltonian}: a 2DES interacting with a polariton BEC. We are interested in the phase transitions that are possible in this system. We find that the polariton BEC can undergo a phase transition into a supersolid, a superfluid with a spatially ordered structure similar to a crystal. At the same time, there are two closely intertwined instabilities in the 2DES: one towards a CDW phase and the other towards a superconducting phase. All of these transitions are possible due to the softening of the polaritons. We notice the similarity to  strongly correlated electron systems such as high temperature superconductors which exhibit a quantum critical point where many instabilities can occur.

In Section \ref{sec:ResultsSC} we investigate quantitatively the superconducting transition in the 2DES. This transition has been previously investigated~\cite{laussy2010exciton,laussy2012superconductivity,cherotchenko2014superconductivity} without taking into account either the screening effects due to the 2DES or the polariton softening, which we find to be crucial for reaching the strong coupling regime. Moreover, we show in Appendix \ref{app:Superconductivity} that the Fr\"ohlich~\cite{frohlich1952interaction} type potential is not suitable for a reliable calculation of the critical temperature.

In the next subsection we investigate qualitatively the possibility of an unconventional CDW in the 2DES due to the proximity of the BEC to a supersolid phase transition.

\subsection{Superconductivity}\label{sec:ResultsSC}
As mentioned above, as we reach the strong coupling regime the 2DES can become superconducting. Contrary to previous assertions~\cite{laussy2010exciton,laussy2012superconductivity,cherotchenko2014superconductivity} we find that polariton mediated superconductivity is not possible in the presence of electronic screening without taking into account the softening of the polaritons. As the polaritons soften and the polariton BEC approaches the supersolid transition, the electron-polariton interaction greatly increases and the system enters the strong coupling regime. Since materials with lower dielectric constants are more suitable for reaching the strong coupling regime (as we will show in Section \ref{sec:Materials}), in this section we choose to look at TMD monolayers.

\begin{figure}[t!]
  \centering
   \includegraphics[]{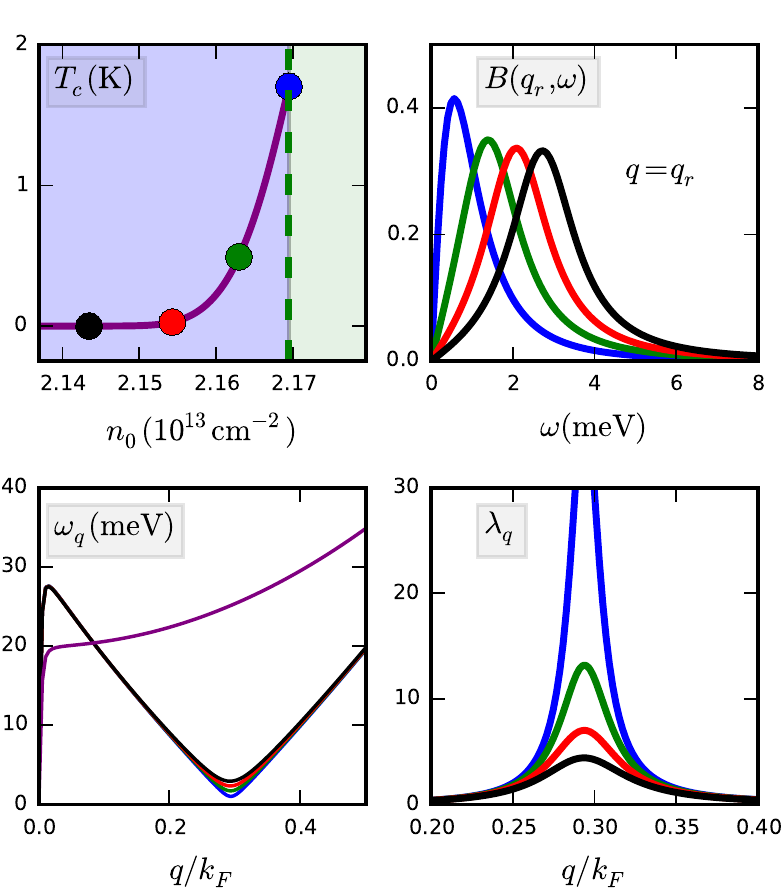}
  \caption[justification=justified,singlelinecheck=false]{Upper-left panel: critical temperature (solid purple line) for a typical TMD monolayer as a function of polariton density $n_0$. The dashed green line is at the critical polariton density $n_c=2.169\times 10^{13}\mathrm{cm}^{-2}$ and can be regarded as dividing the BEC (blue region) and supersolid (green region) phases of the polaritons. Upper-right panel: The polariton spectral function  at the roton-minimum $B(q_r,\omega)$. Lower-left panel: polariton dispersion renormalization (in purple is the bare dispersion). Lower-right panel: $\lambda_q$. The different colors of the lines in the upper-right, lower-left and lower-right panels correspond to different values of $n_0$ denoted by the colored dots in the upper-left panel. Expressed in percentages of $n_c$ these are: $n_0=100\% n_c$(blue), $n_0=99.7\%n_c$(green), $n_0=99.3\%n_c$(red), $n_0=98.8\%n_c$(black).The rest of the parameters are typical TMD monolayers parameters:  $d=a_B=1\mathrm{nm}$, $L=1.5a_B=1.5\mathrm{nm}$, $g_0=10\sqrt{4}\mathrm{meV}$ (4 exciton layers are used), $\epsilon=4 \epsilon_0$,  $m_e=m_h=0.2m_0$, $n_e =10^{13}\mathrm{cm}^{-2}$, $U=0.12\mu\mathrm{eV} \mu \mathrm{m}^2$.}
   \label{fig:MainFigure}
\end{figure}

In the 2D polariton-electron system that we consider, the biggest uncertainty originates from the polariton-polariton repulsion. We note however that the strength of this interaction can be tuned so as to reach a parameter range where our analysis is justified. The other unknown is the broadening of the polariton dispersion around $q_r$ due  to the impurities in the system, which in turn determines the lowest possible energy scale of polaritons at the roton minimum. However, we found out that the polaritons at $q_r$, which contribute most to superconductivity, have an effective mass that is roughly $2$ orders of magnitude lighter than the bare exciton mass (for the parameters used in Figure \ref{fig:MainFigure} and $n_0=n_c$). As a consequence, polaritons at the roton minimum are relatively robust against broadening by static disorder.

The quantity that we are most interested in here is the critical temperature of the superconductor. In Appendix B we provide a short review on how to correctly calculate the critical temperature of a superconductor.

To calculate the superconducting critical temperature we use the modified McMillan formula~\cite{mcmillan1968transition,allen1975transition}:
\bea
k_BT_c &=&\frac{ f_1 f_2 \omega_{log}}{1.2} \exp \left[ - \frac{1.04(1+\lambda) }{\lambda(1-0.62 \mu^*) -\mu^*} \right],\nonumber\\
\mu^* &=& \frac{\mu}{1 + \mu \ln (\varepsilon_F/\hbar \omega_D)}. 
\eea
This formula is only meaningful if the exponent is negative, and is roughly valid for $\lambda < 10$. In the above $\omega_{log}=\exp \left[  \langle \ln(\omega) \rangle \right] $,  (the average is taken with respect to the weight function $\alpha^2(\omega) F(\omega)/\omega$). The screened Coulomb repulsion $\mu$ between electrons averaged over the Fermi surface, is given by:
\bea
\mu = \sum_{k,k'} \frac{V_C (k-k')}{\epsilon(k-k')} \delta(\varepsilon_k) \delta(\varepsilon_k') /N(0) .  
\eea
The correction factors $f_1,f_2$ are given in Appendix \ref{app:Superconductivity}.

In order to calculate the critical temperature we need to know the strength of the exciton-exciton interaction $U(q)$. Given that the exciton-exciton interaction is generally repulsive it has the effect of pushing up the polariton dispersion, stiffening the polaritons. In contrast, the electrons mediate an attractive interaction between polaritons leading to softening. The momentum dependence of the polariton-polariton interaction around $q_r$ has not been investigated, since experiments have been limited to small momentum values of the order of the photon momentum. Theoretical calculations~\cite{ciuti1998role,byrnes2010mott,byrnes2014effective}seem to suggest that in our system the exciton-exciton interaction at large momenta $q\approx 1/a_B$ is about an order of magnitude smaller that the interaction at $q=0$ and might even be attractive. 

Fortunately, since we have a highly tunable system, the highest critical temperatures that we can obtain do not depend strongly on either the strength or the q-space dependence of $U(q)$, as long as the polaritons can soften at some momentum $q$ (i.e. there is a $q$ such that $U(q) + \chi(q) V_X^2(q)<0$). Therefore, given the uncertainty, we choose $U=0.12\mu \mathrm{eV} \mu \mathrm{m}^2$ for our numerical simulations but emphasize that similar results can be obtained by tuning other parameters as long as $U< 0.5 \mu \mathrm{eV} \mu \mathrm{m}^2 $. Furthermore, we reemphasize that tuning the polariton-polariton interaction using Feshbach resonances has been proposed and demonstrated~\cite{carusotto2010feshbach,takemura2014polaritonic}.

In the upper-left panel of Figure \ref{fig:MainFigure} we plot
in solid purple the critical temperatures that can be achieved in a
typical TMD monolayer by tuning the polariton density $n_0$ in a
system with $4$ exciton layers (which have the effect of doubling
$g_0$ compared to the initial value). According to our mean-field calculation in the blue region the 2DES should become superconducting whereas we cannot apply our theory in the green region due to the breakdown of the Bogolyubov approximation. The dashed green line dividing the two regions in phase space is at the critical
polariton density $n_c=2.169\times 10^{13}\mathrm{cm}^{-2}$. At this point $\lambda \approx 1.5$, the condensate depletion is less than $ 5 \%$ and the roton minimum is approximately at $1\mathrm{meV}$. 

In the other three panels we plot the polariton dispersion (lower-left), the polariton spectral function at $q=q_r$ (upper-right) and $\lambda_q$ (lower-right),
which quantifies the attraction strength between two electrons on the Fermi surface resulting from exchanging a virtual BEC excitation of momentum $q$. The exact definition of $\lambda_q$ can be found in Eq.~(\ref{lambdaq}). The different colors of the lines in these three panels correspond to different values of $n_0$ denoted by the colored dots in the upper-left panel Figure \ref{fig:MainFigure}. The blue line correspond to the case of highest polariton density ($n_0=n_c$) while the black line corresponds to the lowest polariton density ($n_0=98.8\%n_c$).

In order to achieve critical temperatures of a few Kelvins the roton-like minimum needs to be lowered to energies of a few meV. As the polaritons soften, the momentum dependent electron-electron attraction, quantified by $\lambda_q$, develops a strong peak at $q=q_r$. This shows that mainly the soft polaritons are responsible for the superconducting phase transition of the 2DES, because they interact most strongly with the 2DES, for two reasons. First of all the softening of the polaritons results in a depletion of the condensate which leads to an increase in the electron polariton matrix element proportional to $\sqrt{\Omega_q/\omega_q}$. Secondly, the electron-electron attraction mediated by the soft polaritons increases even more because these polaritons are  closer to resonance with the electronic interactions which are confined to an energy layer of width $k_B T_c \ll \omega_D$ around the Fermi energy. Looking at the spectral function of the quasi-particles at the roton minimum, denoted by $B(q_r,\omega)$, we notice that the polaritons at the roton minimum become overdamped, similar to paramagnons near a magnetic instability. At this point McMillan's formula should still be valid but one should be careful about the broad polariton spectral function. We investigate the effects of the finite linewidth of this spectral function in Appendix \ref{app:ManyBody} and show that the electrons remain good quasi-particles and that the broad polariton spectral function will not have a significant impact on the superconducting critical temperature.

Another remarkable feature of the electron-electron attraction mediated by polaritons is that it favors p-wave pairing (or other higher symmetries) to s-wave pairing. This is easily understood if we look at the total electron-electron interaction in real space, which we present in Figure~\ref{fig:AppRealSpace} in the Appendix. Notice that this interaction is formed by a strongly repulsive part at the origin followed by a oscillatory part with the wavelength $2 \pi /q_r$, due to the strongly peaked interaction in momentum space. Because of this shape of the interaction, the s-wave pair will feel the strongly repulsive interaction at the origin, while the p-wave pair will avoid this region due to the Pauli-exclusion principle. In accordance with this simple picture, we find p-wave critical temperatures a few times higher than the s-wave critical temperature. However, since the electrons in the p-wave pair are not time-reversal partners these pairs will be influenced by the disorder in the system. Therefore, an accurate calculation of the p-wave critical temperatures requires an estimation of the randomness in our system. 

Notice the strong dependence of the critical temperature on the polariton
density, which indicates that some fine tuning will be necessary in order to observe the superconducting phase. Fortunately, the polariton density is proportional to the intensity of the laser generating the condensate, and the intensity of a laser can be tuned with extreme accuracy. Laser intensities with less than $0.02\%$ noise have been maintained relatively easily in the context of polaritons by using a feedback loop~\cite{abbaspour2015effect}.  We also emphasize that we tried to be conservative in our choice of bare system parameters. It may for example be possible to obtain higher $T_c$ if the polariton density can be increased further without reaching the Mott transition.

\subsection{Supersolid and Charge Density Waves}\label{sec:ResultsGeneral}
As we approach the regime of strong coupling characterized by a significant softening of the polaritons, the system becomes susceptible to other instabilities, in addition to the superconducting instability. When the polaritons soften to the degree that the polariton dispersion touches zero there will be an instability in the polariton BEC. A transition to a supersolid phase  occurs in the green region in Figure~\ref{fig:MainFigure} since the polariton dispersion touches zero roughly at $n_c\approx2.173\times 10^{13}\mathrm{cm}^{-2}$. Even though such a supersolid instability was proposed for indirect excitons~\cite{shelykh2010rotons,matuszewski2012exciton}, we  remark that this phase transition can be more easily observed in a polariton BEC not least because the realization of an exciton BEC is still an experimental challenge. In our theoretical framework based on the Bogolyubov approximation, the onset of this instability can be observed as a dramatic increase in the BEC depletion as the polariton dispersion approaches zero. Since our analysis is only valid when the condensate depletion is small from now on we assume that the polariton dispersion never touches zero. 

As the polariton system approaches its BEC instability to a supersolid, the 2DES becomes susceptible to instabilities mediated by the soft polaritons. In the previous section we analyzed the susceptibility of the 2DES towards a superconducting phase. However, the  strongly peaked electron-electron attraction at $q=q_r$, as shown in the bottom-right panel of Figure~\ref{fig:MainFigure} can result in a CDW state. 
 
A CDW order can also appear due to the phase transition
of the BEC into a supersolid. This transition is
analogous to the case of `frozen phonons' which has been proposed as
an explanation for the CDW order in materials such as TMDs~\cite{weber2011electron,calandra2011charge,joe2014emergence}, where a
finite-momentum phonon softening leads to a condensation resulting
in a static distortion in the lattice which in turn leads to a modulation in the electron
density. In our case this corresponds to the
fact that the mean field $\langle b_{q_r} \rangle$ becomes important (the momentum direction should be chosen spontaneously), which would require
a careful extension of our Bogolyubov approximation scheme.

Remarkably, in contrast to the conventional behavior based on
nesting features in the electron band structure,  this type of
singularity is not originating from the electronic response function
but due to a singular behavior in the electron-polariton interaction
at some wavevector. In both cases mentioned above one
can tune the wavevector where the polariton dispersion touches zero
and therefore can tune the nesting wavevector $q_r$.

We reemphasize that in order to observe a superconducting or CDW phase transition in the 2DES, the polaritons have to soften and the BEC has to be close to the super solid phase transition. More generally, despite the differences in structure and phenomenology, the phase diagram of many unconventional superconductors exhibit the common trait that superconductivity resides near the boundary of another symmetry breaking phase. Examples are the superconducting phases appearing at magnetic quantum phase transtions as found in many of the Ce-based heavy fermion compounds such as $\mathrm{CeIn}_3$~\cite{mathur1998magnetically,monthoux1999p,monthoux2001magnetically,monthoux2005magnetic}, or in iron pnictides  accompanying spin density wave states~\cite{canfield2010feas}, as well as magnetic, stripe and nematic orders discussed for copper oxides~\cite{tranquada1995evidence}. Another example, similar in some respects to our system, is the TMD family, where charge density wave order competes with superconductivity and this feature has been attributed to a softening of the finite momentum phonon modes~\cite{weber2011electron,calandra2011charge,joe2014emergence}.


\section{Materials suitable for reaching the strong coupling regime}\label{sec:Materials}
In this section we do a systematic analysis of the materials most suitable for reaching the strong coupling regime where the 2DES becomes unstable towards a new phase. We find that materials with low dielectric constants are most suitable and conclude that semiconducting TMD monolayers are good candidates for observing polariton mediated superconductivity. 

At a first sight, it may seem that there are many parameters that influence the electron-polariton interaction mediated instabilities that can change from material to material: $\epsilon$, $m_e$, $a_B$, $d$, $L$, $g_0$, $k_F$. However, these parameters are typically not independent. In fact, we argue that all of these material parameters scale with the dielectric constant $\epsilon$ as shown in Table~\ref{tab:title}. 
\vspace{.15in}
\begin{table}[h!]
\caption {Parameter dependence on the dielectric constant $\epsilon$} \label{tab:title}
\begin{tabular}{| l  l  l l |}
\hline\hline
$m_e\quad$ &$\propto$&$\quad$ &$\quad\epsilon^{-1}$ \\
$a_B\quad$ &$\propto \epsilon/m_e  $&$\propto$ & $\quad\epsilon^2 $\\
$d\quad$ &$\propto a_B $&$\propto$ & $\quad \epsilon^2$\\
$L \quad$&$\propto a_B  $&$\propto$ & $\quad \epsilon^2  $\\
$g_0\quad$ &$\propto a_B^{-1} $ &$\propto$ & $\quad \epsilon^{-2}$\\
$k_F\quad$ &$\propto L^{-1}  $&$\propto$ & $\quad\epsilon^{-2}$ \\
$n_0 \quad$&$\propto a_B^{-2} $&$\propto$ & $\quad \epsilon^{-4}$\\
\hline
\end{tabular}
\end{table}
\vspace{.15in}

First, we emphasize that the mass dependence in the  first row is
more complicated and  one may even treat $m_e$ as an independent
parameter as well. For the dipole dependence $d$ we assumed that,
regardless of the mechanism, the induced dipole can be of the order
of the Bohr radius but not larger. Similarly, we assumed that the
distance $L$ between the 2DES and polariton planes cannot be smaller
than the exciton Bohr radius, to avoid tunneling between the two
planes. We also assumed that the light-matter coupling is
proportional to $a_B^{-1}$. It turns out that because
of the large momentum cutoff due to the finite distance $L$ between
the 2DES and polariton planes we get better results with decreasing
$k_F$. However, we cannot lower $k_F$ arbitrarily since we still
need RPA to be valid. Finally, in the last row, we assumed that the
maximum value of $n_0 a_B^2$ is a material independent constant
since it is set by phase space filling~\cite{rochat2000excitonic}.

The strong coupling regime can be characterized by a large EPC constant $\lambda$ and a small Coulomb repulsion constant $\mu$. Therefore we need to investigate the dependence of these parameters on the dielectric constant. Introducing the variable $u\equiv q/2k_F$ and the material independent constants $\bar{L}=2 k_F L\propto \epsilon^0$ and $\bar{k}_{TF}=k_{TF}/(2 k_F)\propto \epsilon^0$:
\bea \label{lam}
\lambda(\epsilon) &\propto&\epsilon^{-3} \int_0^{1}\di u\frac{e^{-2 u \bar{L} }}{ \sqrt{1-u^2} (1+\bar{k}_{TF}/u)^{2} }\cdot \frac{\Omega_{2k_Fu}}{\omega_{2k_F u}^2},  \nonumber \\
\mu (\epsilon) &=&  \epsilon^0 \int_0^{1}\di u\frac{1}{ \sqrt{1-u^2} (1+\bar{k}_{TF}/u) }.
\eea

Looking at the above expressions it is clear that $\mu$ remains roughly constant from material to material which is not unexpected (if we chose $m_e$ as an independent parameter we would then get some variance in $\mu$ for different materials). However, since the electron-polariton interaction is retarded the relevant constant $\mu^*$ given in Eq.~ (\ref{McMillan}) indicates that we can decrease the effective electron-electron repulsion by choosing materials with large Fermi energies and small Debye frequency.

We also see that $\lambda$ depends both on the dielectric constant $\epsilon$ and on the bare and renormalized polariton energies. All of these three parameters can be tuned independently to some extent through various methods.  We notice that observation of polariton mediated superconductivity requires materials with smaller dielectric constants, because smaller dielectric constants favor the dipole interaction over the monopole Coulomb repulsion. This explains why $\lambda \propto \epsilon^{-3}$ while $\mu \propto \epsilon^{0}$.

In addition to a small dielectric constant we want a large bare polariton energy and a small renormalised polariton dispersion. The dielectric constant also sets the value of $g_0$ which gives the energy scale of the bare polariton energy. It should be emphasized though that the bare polariton energy can also be tuned through other methods. One method is to tune the detuning between excitons and cavity and make the polaritons more excitonic or more photonic. This will also change the electron-polariton coupling, an effect which is not captured in Eq.~( \ref{lam}). Another method is the use of multiple quantum wells. By using $N$ quantum wells one can increase $g_0$ by $\sqrt{N}$ while leaving the other parameters unchanged.

Another interesting quantity to investigate is what is the largest renormalization that can be obtained as a function of the dielectric constant. Looking at Eq.~( \ref{omegaq}) and setting $U(q)=0$ we see that the largest renormalisation that can be obtained is given by $\Delta \omega = 2 N_0 \chi(q_r) V_X^2(q_r)\propto \epsilon^{-3}$.

To conclude, we note that under the conditions described in the preceding paragraphs, $\lambda(\epsilon) \propto \epsilon ^{-6}$.

\section{Conclusion}\label{sec:Conclusion}

The striking feature of the coupled 2DES-polariton that we analyzed is the rather unusual nature of the long-range boson-fermion interaction which is peaked at a finite wavevector $q_0$. The latter can be tuned by choosing the system parameters and leads to the emergence of a roton-like minimum in polariton dispersion at $q_r \sim q_0 \neq k_F$. While we have primarily focused on the prospects for observation of light-induced superconductivity, a very interesting open question is the competition between superconductivity and polariton supersolid/CDW phases. The boson-fermion system that we consider allows for precise tuning of key parameters and can be realized either in GaAs or TMD based 2DES/microcavity structures.

{ Strictly speaking, the onset of superconductivity in
the 2DES will be due to a Berezinsky-Kosterlitz-Thouless transition
(BKT) transition. On the other hand, in the parameter range we consider, the $T_c$ we estimate using the mean-field approach should be
comparable to the BKT transition temperature.

A very promising extension  of our work is the realization of
photo-induced p-wave pairing of composite fermions in the quantum
Hall regime~\cite{Moore1991}. Unlike the proximity effect due to an
s-wave superconductor, finite-range polariton-mediated attractive
interaction is more likely to be compatible with requirements for
observing fragile fractional quantum Hall states, enabling
edge-state pairing that was proposed as a method to realize
parafermions~\cite{Clarke2013}.

In contrast to phonons, polaritons can be directly monitored by
imaging the cavity output. This provides a unique possibility to
simultaneously monitor changes in transport properties of electrons
and the spatial structure of polaritons, as the strongly coupled
system is driven through an instability. The driven-dissipative
nature of the polariton condensate could also be used to inject
polaritons at a preferred wavevector $\tilde{q}_r$. Since
$\tilde{q}_r$ need not be equal to the roton minimum $q_r$,
externally imposing a spatial structure for the polariton condensate
could alter the competition between superconductivity and CDW
instabilities. Moreover, the competition between the CDW
and superconductivity can be investigated further by imposing a periodic potential on photonic or excitonic degrees of freedom.

Last but not least, the possibility to turn superconductivity in a
semiconductor high mobility electron system on or off using laser
fields could form the basis of a new kind of transistor that is
compatible with the existing semiconductor nanotechnology. A crucial
development for such applications is a substantial increase of
$T_c$, which may be achieved by properly choosing the properties of
the semiconductor host material.}

\section{Acknowledgements}
The Authors acknowledge useful discussions with Sebastian Huber and Alexey Kavokin.


\appendix
\section{Theoretical analysis}
\label{app:ManyBody}
To tackle the many-body problem we use a Green's functions approach. After doing a mean field approximation the electron-polariton Hamiltonian has a similar structure to the well understood electron-phonon Hamiltonian. Therefore, we will use Migdal-Eliashberg theory to analyze this system theoretically. However, some of our results can also be obtained through a more intuitive canonical transformation and therefore we also present this method in Appendix~\ref{app:Canonical}.
\subsection{Migdal-Eliashberg theory}
We start from the initial Hamiltonian in Eq.~(\ref{GeneralH}). When most of the polaritons are in the BEC ground state at $k=0$ we can simplify this Hamiltonian using the Bogolyubov prescription~\cite{pitaevskii2003bose}, which is equivalent to making the following replacement:  $ b_0 = b_0^\dagger = \sqrt{N_0}$ ($N_0$ is the number of polaritons in the condensate). This is followed by a Bogolyubov approximation which consists in ignoring terms of lower order in $N_0$.

The resulting Hamiltonian is:
\bea\label{StartingH}
H &=& H_{0}^{(e)}+H_{0}^{(p)}+H_{I}^{(e-e)}+H_{I}^{(e-p)}+H_{I}^{(p-p)},\nonumber\\
\eea
where all the other terms remain the same as before except for:
\bea
H_I^{(p-p)} &=&N_0  \sum_{k \neq 0} \frac{U(k)}{2} \left( b_k b_{-k} + b^\dagger_k b^\dagger_{-k} +2 b^\dagger_k b_k \right)\nonumber\\
& &+N_0 U(0) \sum_{k \neq 0} b_k^\dagger b_k, \nonumber\\
 H_I^{(e-p)} &=&\sqrt{N_0} \sum_{k,q} V_X(q) c_{k+q}^\dagger c_k \left( b^\dagger_q + b_{-q}  \right).   
 \eea

After the mean-field Bogolyubov approximation the electron-polariton interaction has the same structure as the electron-phonon interaction and therefore we can analyze it in analogy with the Migdal-Eliashberg theory, which is controlled by the small parameter $\hbar \omega_D/\varepsilon_F$, the ratio of the characteristic phonon/electron energy scales.  Although in doing the many body theory we treated all interactions simultaneously in order to avoid double counting, we choose to present our results in a more intuitive order.

We define the bare electron propagator as
\bea
{\cal G}_e^{(0)}(p)= - \langle T_\tau c_k(\tau) c_k^\dagger(0) \rangle=\frac{1}{i\omega_n - \varepsilon_k},  
\eea
where $p=(i \omega_n, k)$. In light of the following analysis we define the bare polariton propagator as
\bea
{\cal G}_{11}^{(0)}(p)= - \langle T_\tau b_k(\tau) b_k^\dagger(0) \rangle = \frac{1}{i\omega_n - \Omega_k}. 
\eea

The analogy with phonons is made by introducing a phonon-like operator and propagator:
\bea\label{phonon}
A_{q}&\equiv&b_{q}+b^\dagger_{-q},\nonumber\\
{\cal D}(q,\tau) &\equiv &- \langle T_\tau A_q(\tau) A_{-q}(0)   \rangle.
\eea
In the absence of interactions we have:
\bea
{\cal D}^{(0)}(p)={\cal G}^{(0)}_{11}(p)+{\cal G}^{(0)}_{11}(-p)=\frac{2 \Omega_k}{(i \omega_n)^2 - \Omega_k^2 }.  
\eea

As mentioned above we will understand the electron-polariton interaction in terms of the Migdal Eliashberg theory of the electron-phonon interaction. To make any progress, we need to make a Born-Oppenheimer approximation, which in the many-body physics language means that we ignore the electron-phonon vertex corrections due to phonons. This approach is justified by Migdal's theorem~\cite{migdal1958interaction}. We give a brief argument, based on a phase space analysis, that summarizes Migdal's theorem as it applies to our system. For a more rigorous proof one should check Refs. \onlinecite{migdal1958interaction,scalapino1969electron}.

As we will show in the following sections most of the polaritons that interact with the electrons on the Fermi surface are found in a narrow energy interval. Therefore, we can associate an energy scale to the polaritons which we will denote by $\omega_D$, in analogy with the Debye energy in the phonon case. We will define this energy quantitatively below but for now we will assume that such an energy scale can be associated to the polaritons and furthermore we make the assumption that $\hbar \omega_D \ll \varepsilon_F$.

\begin{figure}[h!]
  \centering
  \includegraphics[scale=0.36]{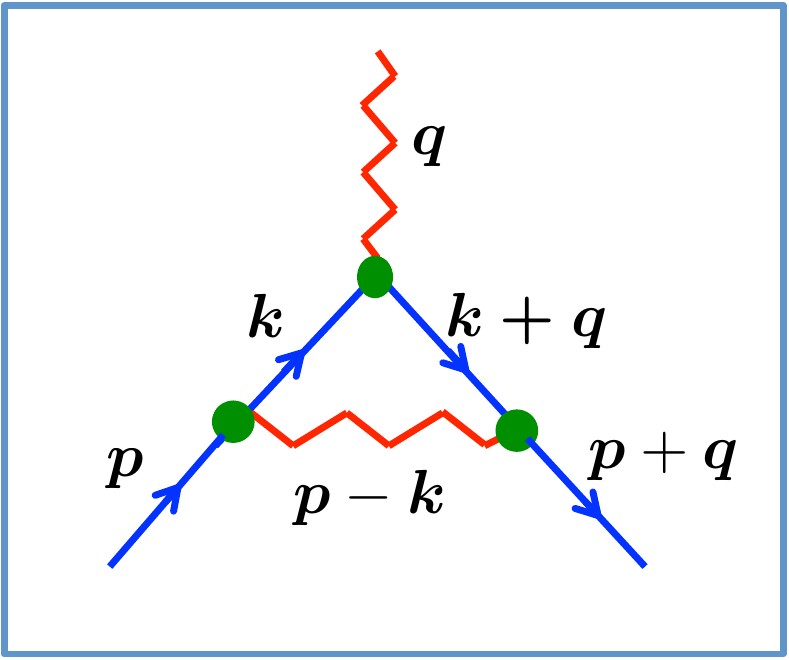}
  \caption{First vertex correction diagram. In blue are the electron propagators while in red are the polariton propagator.}
  \label{fig:FD_2}
\end{figure}

Let us consider the first correction to the electron-polariton vertex, with the corresponding Feynman diagram presented in Figure~\ref{fig:FD_2}. If this correction can be ignored than we can certainly ignore the higher order corrections. Suppose a polariton of momentum $q$ decays and forms an electron-hole pair of momenta $k+q$ and $k$ respectively. This pair will be coherent for a distance of $1/q$. Since the electrons move at roughly the Fermi velocity this gives us a coherence timescale of $1/q v_F$. Only in this timescale electrons can be scattered again. Since the polaritons take much longer to respond we expect this vertex correction to be of order $\omega_D/ q v_F$.  The average momentum of the phonon is of the order of $k_F$ so we expect the vertex correction to depend on the small parameter $\hbar \omega_D/\varepsilon_F$. (If we wish to be more accurate the first vertex correction is of the order of $\lambda \hbar \omega_D/\varepsilon_F$, where $\lambda$ is the electron-polariton coupling constant to be defined below). Thus, we can safely conclude that vertex corrections can be ignored.

The only nontrivial effect of many-body interactions between electrons is the renormalisation of interaction between quasiparticles, i.e. screening~\cite{mahan2000many}. In the following we will explore the screening of both the electron-electron and the electron-polariton interactions in the random phase approximation (RPA). This effect appears as a renormalisation of the photon propagator, where, according to the RPA approximation, the photon proper self-energy is approximated by the simplest polarisation bubble.

In the RPA framework, the screened electron-electron and electron-polariton interaction is expressed in terms of the  dielectric function $\varepsilon(q,i\omega_n)$, which has the following analytical form:
\bea\label{responsefunction}
\epsilon(k,i \omega_n) &=& 1- V_C(k) \chi_0(k,i\omega_n)\nonumber ,\\
\chi_0(k,i\omega_n)&=& \frac{1}{\beta} \sum_{k,ik_n}  {\cal G}^{(0)}(k+q, i\omega_n +i k_n) {\cal G}^{(0)} (k, i k_n)\nonumber \\
&=& \sum_{k} \frac{f(\varepsilon_k) - f(\varepsilon_{k+q})}{i \omega_n + \varepsilon_k - \varepsilon_{k+q}} ,\nonumber \\
\chi (k,i\omega_n) &=& \frac{\chi_0(k,i\omega_n)}{\epsilon(k,i \omega_n)}.
 \eea
In the above, $f(\varepsilon)$ is the Fermi distribution function. The polarisation bubble $\chi_0$ is the Lindhard function and denotes the linear response to a perturbation when electron-electron interactions are neglected. In this case the perturbation disturbs the electron system and creates electron-hole pairs and thus polarising the system. In contrast. the screened polarisation bubble $\chi$ is the response function when electron-electron interactions have been taken into account in the RPA.

The dielectric function for a $2D$ system at zero temperature has been calculated for the first time in Ref. \onlinecite{stern1967polarizability}. The poles of the dielectric function give us the collective excitations of the electron system, the plasmons. Unless otherwise indicated, for the rest of the paper we will take the static limit, (a.k.a. the Thomas-Fermi limit) because it is easier to handle. This limit is accurate as long as the frequencies involved are much smaller than the plasma frequency (in 2D, the only regime where this limit is not satisfied is at very small momenta). In the static limit we have:
\bea
\epsilon(k,\omega) \approx \epsilon(k,0)= \frac{1}{1+\frac{k_{TF}}{k}},  
\eea
where $k_{TF}=m_e e^2/(2 \pi \epsilon \hbar^2)=2/a_B$ is the Thomas-Fermi wavevector.

In terms of this dielectric function, the screened electron-electron and electron-polariton interaction is given by:
\bea
\tilde{V}_C(k) = \frac{V_C(k)}{\epsilon(k)},\nonumber \\
\tilde{V}_X(k) = \frac{V_X(k)}{\epsilon(k)}. 
\eea

The dilute Bose-condensed gas is another one of the few many body systems that are well understood. In this case the small parameter which allows a controlled expansion is given by $n_0 a^2$, where $n_0$ is the polariton density and $a$ is the scattering length of the bosonic repulsion. The field theoretical treatment of the problem was first developed by Beliaev~\cite{beliaev1958application}. A more accessible exposition of this formalism is presented in Ref.~\onlinecite{shi1998finite}.

In addition to the normal polariton propagator introduced above, when interactions are turned on there is an additional anomalous propagator that must be considered. Together, these propagators satisfy the Dyson-Beliaev equations. These equations can be written more compactly by introducing two additional propagators, which are, however, not independent of these two. In the end, the $4$ propagators that need to be considered are:
\bea
{\cal G}_{11}(k,\tau) &=&-  \langle b_k(\tau) b_k^\dagger(0) \rangle,\nonumber\\
{\cal G}_{12}(k,\tau) &=&-   \langle b_{k}(\tau) b_{+k}(0)\rangle, \nonumber \\
{\cal G}_{21}(k,\tau) &=&-  \langle b_{-k}^\dagger(\tau) b_{k}^\dagger(0)\rangle, \nonumber\\
{\cal G}_{22}(k,\tau) &=&- \langle b_{-k}^\dagger(\tau) b_{-k}(0) \rangle   .
\eea
These propagators and the associated self energies are not independent but satisfy the following identities:
\bea
{\cal G}_{22}(p)={\cal G}_{11}(-p),\quad {\cal G}_{12}(p)= {\cal G}_{21}(-p),  \nonumber\\
\Sigma_{22}(p)=\Sigma_{11}(-p), \quad \Sigma_{12}(p) = \Sigma_{21}(-p).  .
\eea

Solving the Dyson-Beliaev equations in the Bogolyubov approximation and making the RPA approximation we obtain the renormalised propagators:
\bea
{\cal G}_{11}(p)&=& \frac{i\omega_n + \Omega_k+\Sigma_{12}(p)   }{(i \omega_n)^2  - \Omega_k^2 - 2  \Omega_k \Sigma_{12}(p) }, \nonumber \\
{\cal G}_{12}(p)&=& \frac{-\Sigma_{12}(p) }{(i \omega_n)^2  - \Omega_k^2 - 2  \Omega_k\Sigma_{12}(p) }, \nonumber\\
\Sigma_{12}(p)&=&N_0 \left[U(k)+\chi(k,i\omega_n) V_X^2(k) \right].  
\eea
The dispersion of the renormalised polaritons, denoted by $\omega_k$, is obtained from the zeros of the real part of the denominator, therefore:
\bea
\omega_k = \sqrt{\Omega_k^2+ 2 \Omega_k N_0 \left[U(k)+\chi(k) V_X^2(k)  \right] },  
\eea
where we made the approximation $\Re[\chi(k,i\omega_n)] \approx \chi(k)$. Notice that the effect of electron-polariton interaction is to renormalise the polariton-polariton interaction. Because this is a second order interaction the effect is proportional to the square of the bare electron-polariton interaction and the response function $\chi$ which also contains the effects due to screening.

The polariton spectral function linewidth comes from the imaginary part of the response function, which is zero in the static limit, so we need to use the frequency dependent response function to calculate the polariton linewidth. Using $ \Omega_k \left| \Im[ \Sigma_{12}(k,\omega)] \right|=\omega_k \gamma_k $ ($\gamma_k$ is the polariton linewidth) we obtain, in the limit $\left| \Im[\epsilon(k,\omega) ]\right| \ll \left| \Re[\epsilon(k,\omega)] \right|$, the following expression for the polariton linewidth:
\bea
\gamma_q &=& 2  \frac{N_0 V_X^2(q)}{\epsilon^2(q)} \frac{\Omega_q}{\omega_q} \Im\left[ \chi _0 (q,\omega_q) \right]\nonumber  \\
&=&2\pi \frac{N_0 V_X^2(q)}{\epsilon^2(q)} \frac{\Omega_q}{\omega_q} \sum_k \left( f(\varepsilon_k) - f(\varepsilon_{k+q}) \right) \nonumber \\
& &\cdot \delta(\omega_q - \varepsilon_k + \varepsilon_{k+q}). 
\eea

As mentioned above, the polarisation bubble has been evaluated exactly in Ref. \onlinecite{stern1967polarizability} but to get a simpler analytical formula we make the following approximation. At low temperatures the Fermi factors restrict the $k$ integration to a narrow region about the Fermi surface of width $\omega_q$, so for reasonably well behaved Fermi surfaces we can replace these factors with $\omega_q \delta(\varepsilon_k)$ to obtain:
\bea
\gamma_q &\approx& 2\pi \frac{N_0 V_X^2(q)}{\epsilon^2(q)} \frac{\Omega_q}{\omega_q} \omega_q \sum_{k}  \delta(\varepsilon_k) \delta(\varepsilon_{k+q}).  
\eea
We remark that the same result can be obtained using Fermi's golden rule if we consider the renormalised electron polariton interactions to be given by $\tilde{M}_q=\sqrt{\frac{N_0 V_X^2(q)}{\epsilon^2(q)} \frac{\Omega_q}{\omega_q}}$ as shown in Refs. \onlinecite{allen1972neutron,allen1974effect}. We will find out below that this is indeed the proper renormalised electron-polariton interaction.

Evaluating the sum we obtain:
\bea
\frac{\gamma_q}{\omega_q}\approx \frac{ \tilde{M}^2_q N (0)}{\varepsilon_F} \frac{2 k_F}{q \sqrt{1-(q/2 k_F)^2}},  
\eea
where $N(0)$ is the electron density of states at the the Fermi surface. In order to have long lived quasi-particles we need to satisfy $\gamma_q/\omega_q \ll 1$.

Having discussed the condensate properties in a field theoretical formalism, we investigate the effect of the condensate on the electrons and we see that polariton excitations can mediate an attractive interaction between electrons.

In the presence of interactions, the phonon like propagator introduced in Eq.~( \ref{phonon}) has the form:
\bea
{\cal D}(p)&=&{\cal G}_{11}(p)+{\cal G}_{22}(p)+{\cal G}_{12}(p)+{\cal G}_{21}(p) \nonumber\\
&=& \frac{2  \Omega_k}{\left( i \omega_n\right)^2 -\omega_k^2 - 2 i \gamma_q \omega_q}.  
\eea

Notice that the propagator depends on both the bare and the renormalised polariton spectrum. As usual~\cite{mahan2000many}, we define a \emph{reduced} propagator $\bar{\cal D} $ which corresponds to the propagation of the new polariton quasi-particles and therefore depends only on the renormalized quasi-particles' spectrum:
\bea
{\cal D}(p)&=& \frac{\Omega_k}{\omega_k} \bar{\cal D}(p).  
\eea

The polariton mediated electron-electron attraction can be expressed in terms of this propagator:
\bea
V_{e-e}^{(eff)}(p)&=&\frac{V_C(k)}{\epsilon(k)} + \frac{N_0 V_X^2(k)} {\epsilon^2(k)} {\cal D}(p) \nonumber\\
&=&\frac{V_C(k)}{\epsilon(k)} + \tilde{M}_k^2 \bar{\cal{D}}(p) ,
\eea
where we have introduced the renormalised electron-polariton matrix element
\bea
\tilde{M}_q=\sqrt{\frac{N_0 V_X^2(q)}{\epsilon^2(q)} \frac{\Omega_q}{\omega_q}}.  
\eea

Notice that the term $\Omega_k/\omega_k$ from the initial propagator ${\cal D}(p)$ has been absorbed in the electron-polariton matrix element. In the previous section we noticed that this is necessary in order to obtain the same polariton linewidth as the one calculated using Fermi's golden rule and now we have seen why.

We also investigate the electron self energy acquired through interactions with polaritons. This contribution is small but its derivative with respect to energy is large within $\omega_D$ of the Fermi surface. Therefore it will strongly affect the electrons within $\omega_D$ of the Fermi surface. The main effects are a renormalised mass and a finite quasi particle linewidth.

The contribution of the polaritons to the electron self-energy has the following analytical form:
\bea
\Sigma(k,i \omega_n) &=& - \frac{1}{\beta} \sum_{i q_n}\int \frac{\di ^2 q}{(2 \pi)^2} \tilde{M}_q \bar{{\cal D}}(q,i q_n)\nonumber\\
& &\cdot {\cal G}^{(0)}(k+q,i\omega_n + i q_n).  
\eea

The resulting effects are conventionally expressed in terms of the Eliashberg function $\alpha^2 F(\omega)$. This function is closely related to the polariton density of states:
\bea
F(\omega) = \sum_q \delta(\omega - \omega_q).  
\eea
However, the connection is somewhat obscured in the usual definition:
\bea
\alpha^2 F (\omega) = \sum_{k,k'} \left| \tilde{M}_{k-k'} \right|^2 \delta(\omega- \omega_{k-k'}) \delta (\varepsilon_k) \delta (\varepsilon_k') /N(0).  \nonumber\\
\eea
The above function can be expressed in terms of the previously investigated polariton linewidth $\gamma_q$:
\bea
\alpha^2 F (\omega) = \frac{2}{\pi} N(0) \omega \sum_q \gamma_q \delta(\omega- \omega_q). 
\eea

Most properties of the electron-polariton interaction can be expressed in terms of the EPC (electron-polariton coupling in our case) constant $\lambda$ and averages $\langle \omega^n \rangle$, where $n$ are integers and the average is taken with respect to the weight function $\alpha^2 F(\omega)$. For example, the Debye frequency that we defined above can be expressed quantitatively as:
\bea
\omega_D  \equiv \langle \omega \rangle =    2 \int_{0}^{\infty} \di \omega\ \alpha^2 F(\omega)/ \lambda.  
\eea

The definition of the EPC constant is:
\bea
\lambda = 2 \int \frac{\di \omega \ \alpha^2 F(\omega)}{\omega}.  
\eea
This constant can also be expressed in terms of the $\lambda_q$ which makes explicit the contribution of polaritons with different momenta:
\bea\label{lambdaq}
\lambda=\frac{1}{N} \sum_q \lambda_q = \frac{1}{\pi N(0)} \sum_q \frac{\gamma_q}{\omega_q^2},  
\eea
where $N$ is the total number of electrons, which in $2D$ is given by $N=N(0) \varepsilon_F$.

Returning to the electron self energy due to interactions with polaritons, the real part results in a mass renormalisation of the electron quasi particle given by:
\bea\label{effectivemass}
m^*_e&=&m_e(1+\lambda)\to \tilde{\varepsilon}_k =\frac{\varepsilon_k}{1+\lambda}.
\eea

The imaginary part of the self energy gives the electron quasi particle linewidth $\Gamma$. At zero temperature~\cite{grimvall1981electron}:
\bea \label{ELife}
\Gamma (\omega) =\pi \int_0^\omega \di \omega' \alpha^2 F(\omega').
\eea

Clearly, the electron quasi-particles with energies close to the polariton energy scale $\omega_D$ will be short lived because electrons will be able to lose their energy to excite polaritons. For these electrons, the quasi particle picture fails. However, in our system, we have the following energy scale $k_B T_c \ll \omega_D$. Therefore, we expect that the superconducting electrons will not be affected by dissipation due to polaritons, so we can still use the quasi particle picture.

\subsubsection{Finite polariton spectral function linewidth}
In the previous discussion we have treated the polaritons as perfect quasi-particles. When the linewidth of the polariton spectral function becomes significant the Eliashberg function needs to be modified and is broadened:
\bea
\alpha^2 F (\omega) &=& \sum_{k,k'} \left| \tilde{M}_{k-k'} \right|^2 B(k-k',\omega) \delta (\varepsilon_k) \delta (\varepsilon_k') /N(0), \nonumber\\
B(q,\omega) &=& \frac{1}{\pi} \Im [D(q,\omega)] = \frac{1}{\pi} \Im \left[ \frac {\Omega_q}{\omega_q} \frac{2 \omega_q}{\omega ^2 - \omega_q^2 - 2 i \gamma_q  \omega_q}\right] ,\nonumber\\
\eea
where the Lorentzian $B(q,\omega)$ is the polariton spectral function. We are interested in how the finite-polariton lie width will influence the superconducting properties of the electron system.

The first question that we need to ask is whether the electrons remain good quasi-particles. As shown in Refs. \onlinecite{allen1972neutron,allen1974effect}, when the finite polariton - linewidth is included the electron lifetime scales as $\varepsilon^2$ close to the Fermi surface. Therefore electrons close to the Fermi surface are well-defined quasi-particles. In all our numerical simulations we checked that electrons are well defined quasiparticles in a shell of the order of $k_B T_c$, where $T_c$ is the superconducting critical temperature.

According to McMillan's formula in Eq.~(\ref{McMillan}) the superconducting critical temperature can be expressed in terms of 4 constants: $\mu^*$ , $\lambda$ , $\omega_{log}=\exp \left[  \langle \ln(\omega) \rangle \right] $, $\bar{\omega}_2=\langle \omega^2 \rangle ^{1/2}$ (the averages are taken with respect to the weight function $\alpha^2(\omega) F(\omega)/\omega$). Only the last 3 constants will be affected by the broadening of the Eliashberg function. Furthermore, for large $\lambda$ the critical temperature $T_c\propto \sqrt{\lambda} \bar{\omega}_2$, therefore we will only investigate how these constants are modified.

We can rewrite $\lambda$ as:
\bea
\lambda = 2 \sum_{k,q} \left| \tilde{M}_q \right| ^2 \delta(\varepsilon_k) \delta(\varepsilon_{k+q}) \frac{1}{N(0)} \int_0^\infty \di \omega \frac{B(q,\omega)}{\omega}.  \nonumber\\
\eea
However, by definition:
\bea
\bar{D}(q,\omega) = \int_{-\infty}^{\infty} \di \omega' \frac{B(q,\omega')}{\omega' - \omega +i \delta} .  
\eea
Using the oddness of $B(q,\omega)$ we obtain:
\bea
\int_0^\infty \di \omega \frac{B(q,\omega)}{\omega} = -\frac{D(q,0)}{2} = \frac{\omega_q}{\omega_q^2(0)},  
\eea
where $\omega_q(0)$ denotes renormalised polariton energy obtained by using the static (Thomas-Fermi) dielectric function. Thus, when the finite polariton lifetime is taken into account, $\lambda$ should be calculated using the polariton energies obtained from the static limit for the polarisation bubble (which is actually what we already did to simplify calculations).

We now investigate the effect of the broadened Eliashberg function on $\lambda \langle \omega^2 \rangle$:
\bea
\lambda \langle \omega^2 \rangle &=& 2 \sum_{k,q} \left| \tilde{M}_q \right| ^2 \delta(\varepsilon_k) \delta(\varepsilon_{k+q}) \frac{1}{N(0)} \int_0^\infty \di \omega \omega B(q,\omega)\nonumber \\
&=&2 \sum_{k,q} \left| \tilde{M}_q \right| ^2 \omega_q \delta(\varepsilon_k) \delta(\varepsilon_{k+q}) \frac{1}{N(0)},   
\eea
where we used the well known~\cite{allen1974effect} sum rule $\omega_q = \int_0^\infty \di \omega\ \omega B(q,\omega)$. Therefore $\lambda \langle \omega^2 \rangle$ is not affected at all by the broadening of the Eliashberg function.

In conclusion we can safely neglect the effect of the polariton linewidth on the superconducting properties of the 2DES.

\subsection{Canonical transformation}\label{app:Canonical}

In this subsection we will show how some of the above results can be obtained in a Hamiltonian formalism using canonical transformations. This approach might be preferred for it's simplicity and because it yields information about the quasi-particle wavefunction.

In order to obtain the same results as in the diagrammatic approach we must first better understand the approximations that we made in the diagramattic approach and make the same approximations in this context. Notice that our choice of the electron and polariton self-energies implies that we treat the electron-polariton interaction up to second order. This approximation was justified by Migdal's theorem. However, the polariton-polariton interactions are treated exactly. In contrast the electron-electron interactions are treated perturbatively in the RPA approximation, which is an infinite sum containing terms of all perturbative orders.

Finally, McMillan's formula is based on the approximation that the electrons that participate in superconductivity are in a very thin layer around the Fermi surface much smaller than the polariton energy scale which allows one to approximate the electrons as living on the Fermi surface. We will need this approximation to avoid complications.

As before we start from the Hamiltonian in Eq.~(\ref{StartingH}) which has been obtained in the Bogolyubov approximation:
\bea\label{NewStartingH}
H &=& H_{0}^{(e)}+H_{0}^{(p)}+H_{I}^{(e-e)}+H_{I}^{(e-p)}+H_{I}^{(p-p)}, \nonumber\\
\eea
where, as before:
\bea
H_I^{(p-p)} &=&N_0  \sum_{k \neq 0} \frac{U(k)}{2} \left( b_k b_{-k} + b^\dagger_k b^\dagger_{-k} +2 b^\dagger_k b_k \right)\nonumber\\
& &+N_0 U(0) \sum_{k \neq 0} b_k^\dagger b_k, \nonumber\\
 H_I^{(e-p)} &=&\sum_{k,q} M_q c_{k+q}^\dagger c_k \left( b^\dagger_q + b_{-q}  \right).  
 \eea
In the above $M_q=\sqrt{N_0}  V_X(q) $.

We will first show how the one can investigate the effect of electrons on polaritons by tracing out the electrons and then we will show how one can investigate the electronic system by tracing out the polaritons.

\subsubsection{Tracing out the electrons}
We first wish to investigate the effect of electrons on the polaritons. The effect of electrons on polaritons is to induce an effective attraction between polaritons.

To get the same results as in the diagrammatic approach we must treat the electron-polariton interaction to second order in perturbation theory, however, the interaction must contain the screening effects due to electron-electron interactions. Intuitively, a polariton of momentum $k$ creates a potential $V_X(k)$ in the 2DES, and the 2DES responds by creating a charge density $\delta n (k) = \chi(k) V_X(k)$ where $\chi(q)$ is the response function first introduced in Eq.~(\ref{responsefunction}). This charge density will create a potential $V_X(k)$ at the polariton condensate, thus resulting in an attraction $\chi(k) V^2_X(k)$ between a polariton of momentum $k$ and a polariton in the condensate. After tracing out the electrons and rearranging terms we obtain the following effective polariton Hamiltonian:
\bea
H^{(p)}&=& \sum_{k\neq 0}\left[E_{k} \left( b_k^\dagger b_k +b_{-k}^\dagger b_{-k} \right) +g_k \left( b_k^\dagger b^\dagger_{-k} +b_k b_{-k} \right) \right],\nonumber\\  
\eea
where the sum must be taken over half of $k$-space to avoid double counting. In the above $g_k \equiv N_0 \left[ U(k) + \chi(k) V_X^2(k) \right]$ is the effective interaction between the condensate and a polariton of momentum $k$. Notice that the polariton energy $E_{k} \equiv \Omega_k + g_k$ contains the Hartree-Fock interaction $g_k$.

The above Hamiltonian is quadratic and therefore can be diagonalized by making a canonical transformation (a.k.a. the Bogolyubov transformation) by introducing the new operators:
\bea\label{BogBog}
a_k = u_k b_k + v_{-k}^* b_{-k}^\dagger . 
\eea
This is a canonical transformation if $\left| u_k \right|^2 - \left| v_{-k} \right|^2 = 1 $. Notice that the phases are irrelevant and therefore we choose $u_k , v_k \in \Re$. To diagonalize the Hamiltonian we must choose:
\bea
u_k, v_{-k} = \pm \sqrt{\frac{1}{2}} \sqrt{\frac{E_{k}}{\omega_k}\pm 1},  
\eea
where we introduced the excitation spectrum $\omega_k \equiv \sqrt{E_{k}^2 - g_k^2} = \sqrt{\Omega_k^2 - 2 \Omega_k g_k}$. The new Hamiltonian is diagonal:
\bea
H^{(p)} = \sum_k \omega_k a^\dagger_k a_k.  
\eea
At this point we have found the polariton excitations on top of the condensate. We wish to investigate the interaction between electrons and the new excitations. To do this we express the electron-polariton Hamiltonian $H^{(e-p)}$ in terms of the new excitations:
\bea
H_I^{(e-p)} &=& \sum_{k,q} M_q \sqrt{\frac{\Omega_q}{\omega_q}} c^\dagger_{k+q} c_k (a_q+ a^\dagger_{-q}).  
\eea
We see that the electron-polariton interaction is increased by the factor $\sqrt{\Omega_q/\omega_q} = u_q - v_q$ which, (not surprisingly) is exactly what we found in the diagrammatic approach. 

The reason for the increase of interactions is that the elementary excitations on top of the condensate contain many polaritons, and therefore they interact more strongly with the electrons. To understand this effect better we look at the interaction between the interacting ground state $\ket{G}$ and an excited state of momentum $k$ which we denote by $\ket{E}_k$. Since $\ket{G}$ is defined by $a_k \ket{G}=0$ for all $k$, we can express the interacting ground state in the Fock basis of the initial/bare polaritons. We denote by $\ket{n}_k$ the Fock state containing $n$ bare polaritons of momentum $k$. In this notation:
\bea
\ket{G}&=&\Pi_k \ket{G}_k, \quad \mathrm{where,} \nonumber \\
\ket{G}_k &=& \frac{1}{u_k} \sum_n \left(\frac{-v_k}{u_k}\right)^n \ket{n}_{k}\ket{n}_{-k}. 
\eea 
As before, the product runs over half of momentum space to avoid double counting. Notice that the ground state factorizes between states of opposite momenta due to translational invariance. We mention that the condensate depletion due to bare polariton excitations at momentum $k$ given by $\bra{G} b^\dagger_k b_k \ket{G}=v_k^2$ tells us how many polaritons of momentum $k$ are contained in the ground state on average. One must always make sure that the total condensate depletion $\sum_k v_k^2$ is small compared to the number of polaritons in the ground state $N_0$.  

The elementary excitation of momentum $k$ is easily found: 
\bea
\ket{E}_k&=&a_k \ket{G}_k= \frac{1}{u_k^2} \sum_n \left(\frac{-v_k}{u_k}\right)^n \sqrt{n+1} \ket{n+1}_k \ket{n}_{-k}  \nonumber\\
\eea 
We also express the electron-polariton interaction in the Fock basis: 
\bea
H_I^{(e-p)} &=& \sum_{k,q} c^\dagger_{k+q} c_k  M_q \\
& &\cdot \sum_n  \sqrt{n}\left(\ket{n-1}_q \bra{n}_q + \ket{n}_{-q}\bra{n-1}_{-q}   \right) .\nonumber
\eea
Notice that the interaction strength depends on the number of polaritons at momentum $q$. Since both the ground state and the excited state contain many polaritons when the condensate depletion $v_q^2$ is large, the electrons will be able to scatter excitations at this momentum much more efficiently. To illustrate this we calculate the amplitude for scattering a bogolon of momentum $q$:
\bea
\bra{G}_q H_I^{(e-p)} \ket{E}_q = \sum_{k,q} c^\dagger_{k+q} c_k  M_q (u_q-v_q).
\eea

\subsubsection{Tracing out the polaritons}
We can also start our analysis by tracing out the polaritons. As in the diagrammatic approach we start from the Hamiltonian in Eq.~(\ref{NewStartingH}) which has been obtained in the Bogolyubov approximation. In this section we are interested in the electronic part of the system and therefore we want to trace out the polaritons.  We can achieve this by performing a Schrieffer-Wolff transformation and keeping only second order terms. There are different Schrieffer-Wolff transformations that can be done as we explain in Appendix~\ref{app:Superconductivity}. However, the situation is simplified if we consider scattering only on the Fermi surface (which we also did in the diagrammatic approach by using McMillan's formula). Because this is an energy conserving process it could be measured experimentally and therefore in this case all the different transformations must yield the same second order result.

The Schrieffer-Wolff transformations mentioned above were derived in the context of electron-phonon interactions. In that case  phonon-phonon interactions were neglected because they are negligible. In order to apply these transformations to our system we need get rid of the polariton-polariton interaction which we do with the help of the Bogolyubov transformation introduced previously in Eq.~(\ref{BogBog}). After doing the Bogolyubov transformation we obtain the Hamiltonian:
\bea
H&=& \sum_k \varepsilon_k c^\dagger_k c_k +  \sum_k \tilde{\Omega}_k \tilde{b}^\dagger_k \tilde{b}_k \nonumber + \sum_{k,k',q} V(q) c^\dagger_{k+q} c^\dagger_{k'-q} c_k c_k' \\
& & + \sum_{k,q} \tilde{M}_q c_{k+q}^\dagger c_k \left( \tilde{b}^\dagger_q +\tilde{ b}_{-q}  \right),  
 \eea
where we denote by $\tilde{b}_k$ the destruction operator for the new non-interacting polaritons, $\tilde{\Omega}_k = \sqrt{\Omega_k^2 + 2 \Omega_k N_0 U(k)}$ is the spectrum of the non-interacting polaritons and $\tilde{M}_q=M_q  \sqrt{\Omega_q/\tilde{\Omega}_q}$ is the new electron-polariton interaction matrix element. So far, we made only the Bogolyubov approximation. 

At this point we can trace out the polaritons using a Schrieffer-Wolff transformation and obtain the following effective Hamiltonian:
\bea
H_{e} &=& \sum_k \varepsilon_k c^\dagger_k c_k + \sum_{k,k',q} V(q) c^\dagger_{k+q} c^\dagger_{k'-q} c_k c_k' , \nonumber \\
V(q)&=&V_C(q)- \frac{2 | \tilde{M}_q |^2}{ \tilde{\Omega}_q}.  
\eea

When taking screening into account in the RPA approximation just as in the diagrammatic approach we find that the screened electron-electron interaction is given by:
\bea
\tilde{V}(q) = \frac{V(q)}{1- V(q) \chi_0(q)} .  
\eea
The above expression can be rewritten in a more intuitive form by separating $\tilde{V}(q)$ into a screened Coulomb repulsion and an effective attraction mediated by polaritons:
\bea
\tilde{V}(q) = \frac{V_C(q)}{\epsilon(q)} + \left( \frac{\tilde{M}_q}{\epsilon(q)} \right)^2 \frac{2 }{\tilde{\Omega}_q +2 \chi(q) \tilde{M}^2(q) },  
\eea
where as before we have introduced the dielectric function $\epsilon(q) = 1 - V_C(q) \chi_0(q)$ and the response function $\chi(q) = \chi_0(q) / \epsilon(q)$. The effective electron-electron interaction can be rewritten in a simpler form by introducing the renormalised polariton energies $\omega_q \equiv \sqrt{\Omega_q^2 + 2 N_0 \left[U(q) + \chi(q) V^2_X(q) \right]}$:
\bea
\tilde{V}(q) = \frac{V_C(q)}{\epsilon(q)} + \left( \frac{M_q}{\epsilon(q)} \right)^2\frac{\Omega_q}{\omega_q} \frac{2 }{\omega_q }.  
\eea
This is exactly the result we found using a diagrammatic approach. The second term in the LHS is the effective attraction between electrons when exchanging momentum $q$ through virtual polaritons. Notice that taking the average on the Fermi surface of the first term in $V(q)$ yields the Coulomb repulsion constant $\mu$ (this constant will be defined in Appendix~\ref{app:Superconductivity}). Also, taking the average over the Fermi surface of the second term in $V(q)$ yields the EPC constant $\lambda$. 

\begin{figure}[h!]
     \includegraphics[]{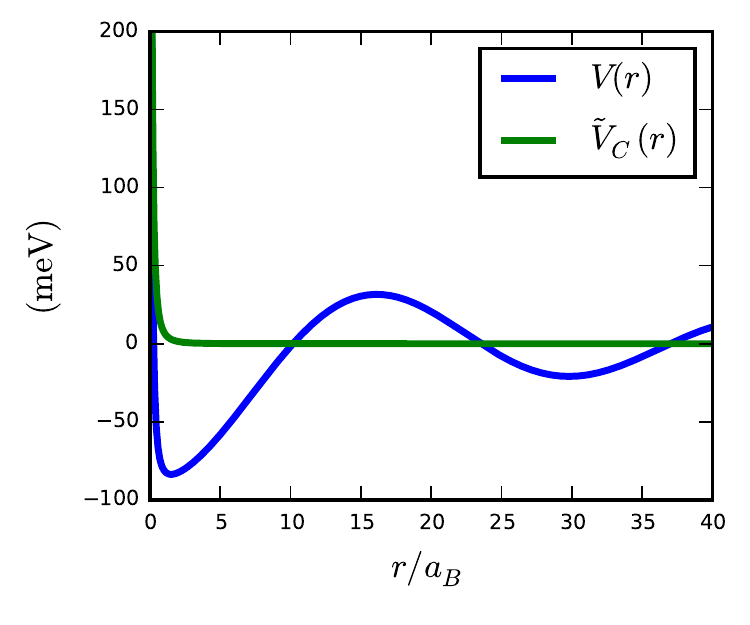}
  \caption{The total interaction $\tilde{V}(r)$ (blue) compared to the screened Coulomb repulsion $\tilde{V}_C(r)$(green) for the parameters used in Figure~\ref{fig:MainFigure} with $n_0=n_c$.}
   \label{fig:AppRealSpace}
\end{figure}

Since it might be useful to have an idea of the shape of the interaction in real space, we look at the Fourier transform of the total interaction compared to the screened Coulomb interaction for some typical TMD monolayer parameters in Figure~\ref{fig:AppRealSpace}. Notice the oscillatory behaviour of the interaction at the wavelength $2 \pi/q_r$ that appears due to the softening of the polaritons. In comparison to the long range attractive interaction, the screened Coulomb interaction looks like a contact interaction. 


\section{Superconductivity}\label{app:Superconductivity}
In this section we briefly review the methods that can be used to calculate the critical temperature of a polariton-mediated superconductor.  This discussion is necessary in the polariton community in order to make clear the connection to superconductivity and to see in which ways our system behaves as a conventional/unconventional superconductor.

We mention that it is notoriously difficult to make quantitative theoretical predictions of the superconducting properties of a new material. However, this limitation is due to the lack of knowledge of the normal state of the material and not due to the accuracy of the BCS theory. In metals, many complications arise which do not concern us, i.e. the choice of the bare pseudopotential describing electron-ion interaction, phonon polarisation vectors, Umklapp processes, distortions of the Fermi surface, etc. However, in our system the normal state can be more readily investigated, because the bosons and the fermions can be separated and investigated separately.  In this respect, this type of superconductivity is most similar to the superconductivity in doped semiconductors, which are, in this sense, the best understood superconductors~\cite{cohen2015superconductivity}.

In previous work on polariton mediated superconductivity, not only renormalisation effects have been ignored, but also the method used to calculate the critical temperature is not valid. Therefore, in this section we review the methods that can be used to make reliable predictions about a new superconductor given that the normal state is known and we point out the reason why the predictions made in previous work cannot be taken seriously.

In Section~\ref{sec:McMillanequation} we present McMillan's equation, which is the simplest method to obtain the critical temperature of a superconductor given some system paramters\footnote{Usually, the McMillan formula is used the other way around: given a critical temperature, system parameters like the EPC constant $\lambda$ can be calculated}. Then, is Section~\ref{sec:BCSequation} we introduce the BCS gap equation, which was initially used by Bardeen, Cooper, and Schrieffer to theoretically explain superconductivity, in order to explain the discrepancy between our results and the results obtained in previous work~\cite{laussy2010exciton} in Section~\ref{sec:CompLaussy}.

\subsection{McMillan equation}\label{sec:McMillanequation}
The state of the art in the theory of superconductivity are the Eliashberg equations, obtained in a Green's Function formalism. In certain limits they can be reduced to a set of two coupled integral equations which must be solved self-consistently. In some limits, which we will present below, these equations can be solved analytically to obtain the critical temperature of the superconductor~\cite{mcmillan1968transition}. Further correction factors can be introduced by fitting to the exact results obtained by numerically solving the Eliashberg equations, to obtain, as shown in Ref. \onlinecite{allen1975transition}:
\bea\label{McMillan}
k_BT_c &=&\frac{ f_1 f_2 \omega_{log}}{1.2} \exp \left[ - \frac{1.04(1+\lambda) }{\lambda(1-0.62 \mu^*) -\mu^*} \right]\nonumber\\
f_1&=&\left[1+\left(\lambda/\Lambda_1\right)^{3/2}\right]^{1/3}, f_2=1+\frac{\left(\bar{\omega}_2/\omega_{log}-1\right)\lambda^2}{\lambda^2+\Lambda_2^2}\nonumber\\
\Lambda_1&=&2.46 (1+3.8\mu^*),\Lambda_2=1.82(1+6.3 \mu^*)(\bar{\omega}_2/\omega_{log})\nonumber\\
\mu^* &=& \frac{\mu}{1 + \mu \ln (\varepsilon_F/\hbar \omega_D)} 
\eea
In the above $\omega_{log}=\exp \left[  \langle \ln(\omega) \rangle \right] $, $\bar{\omega}_2=\langle \omega^2 \rangle ^{1/2}$ (the averages are taken with respect to the weight function $\alpha^2(\omega) F(\omega)/\omega$), and $\mu$, the screened Coulomb repulsion between electrons averaged over the Fermi surface, is given by:
\bea
\mu = \sum_{k,k'} \frac{V_C (k-k')}{\epsilon(k-k')} \delta(\varepsilon_k) \delta(\varepsilon_k') /N(0) .  
\eea

According to Ref. \onlinecite{allen1975transition}, the above formula, known as McMillan's formula, is accurate for  $\mu^*$ ranging between $0 < \mu^* <0.2 $ and $0.3 < \lambda < 10$. Therefore, in order to know the critical temperature of a superconductor, we need to know four material constants $\lambda, \omega_{log},\bar{\omega}_2$ and $\mu^*$. The difficulty lies in accurately determining these parameters. In our case, due to the simplicity of our system, we expect these parameters to be close to the theoretical predictions.

It is useful to express the parameters $\lambda$ and $\mu$ as momentum integrals, rather than frequency integrals. In this case simple analytical expressions can be obtained:
\bea
\lambda &=& \frac{2 N(0)}{\pi k_F }\int_0^{2 k_F}\di q  \frac{\tilde{M}^2(q)}{\omega_q} \left[1-\left(\frac{q}{2k_F}\right)^2\right]^{-1/2},\nonumber \\
\mu &=&  \frac{N(0)}{\pi k_F }\int_0^{2 k_F}\di q  V_C(q) \left[1-\left(\frac{q}{2k_F}\right)^2\right]^{-1/2}.  
\eea

\subsection{Superconducting gap equation}\label{sec:BCSequation}
We wish to compare our method to the method used in previous work on polariton-mediated superconductivity. In order to make this comparison, we show how the superconducting critical temperature can be obtained from a Hamiltonian formalism. Since this section is only meant for comparison we do not consider any renormalisation effects, or the Coulomb repulsion. Therefore, our starting Hamiltonian will be:
\bea
H&=&\sum_k \varepsilon_k c^\dagger_k c_k + \sum_q \omega_q b^\dagger_q b_q \nonumber\\
& &+\sum_{k,q} M_q \left( b_q + b^\dagger_{-q} \right)  c^\dagger_{k+q} c_k .  
\eea

In a Hamiltonian formalism one can obtain an integral equation for the gap function, provided one can trace out the polaritons to obtain an electron-electron attractive interaction between Cooper pairs such that:
\bea
H_{eff} = V(k,k') c^\dagger_{k'}  c^\dagger _{-k'} c_{k} c_{-k}.  
\eea
Supposing that this Hamiltonian can be obtained, then one can apply the methods first introduced by Bardeen, Cooper and Schrieffer~\cite{bardeen1957theory} to obtain the following BCS gap equation (at zero temperature):
\bea\label{GapEquationMomentum}
\Delta(k) = \sum_{k'} V(k,k') \frac{\Delta(k')}{2  \sqrt{\Delta(k')^2 + \varepsilon_{k'}}}.
\eea
Going to a continuum and changing variables from $k$ to $\varepsilon, \theta$ we obtain the above equation in a more convenient form:
\bea\label{GapEquationEnergy}
\Delta(\varepsilon) &=& \int_{-\varepsilon_F}^{\varepsilon_F} \di \varepsilon' \frac{\Delta(\varepsilon')}{2 \sqrt{\Delta(\varepsilon')^2 + \varepsilon'}} V(\varepsilon-\varepsilon') .
\eea

It is not at all obvious how to correctly trace out the polaritons to obtain an electron-electron effective interaction, mainly due to the retarded nature of this interaction  \cite{allen1982solid}. We briefly present three methods and comment on their validity:
\bea
V^F(k,k')&=&\frac{2 | M_q |^2 \omega_q}{\Delta \varepsilon^2 - \omega_q^2} \nonumber, \\
\nonumber V^{BCS}(k,k')&=&\begin{cases} -\frac{2 | M_q |^2}{ \omega_q} &, \mathrm{if}\ |\Delta \varepsilon | < \omega_D \\
0 &, \mathrm{otherwise}
\end{cases}, \\
  V^{S}(k,k')&=&-\frac{2 | M_q |^2 }{|\Delta \varepsilon|+ \omega_q},
\eea
where $q=k-k'$ and $\Delta \varepsilon = \varepsilon_{k'} - \varepsilon_k$.

The first effective potential $V^F$ was initially derived by Fr\"ohlich~\cite{frohlich1952interaction} through a Schrieffer-Wolff transformation to leading order in the electron-phonon coupling. Notice that it has a resonance singularity, which means that at that point higher order terms in the Schrieffer-Wolff transformation become important. However, the singularity is eliminated when performing the (principal value) angular integral which appears from changing variables in going from Eq.~(\ref{GapEquationMomentum}) to Eq.~(\ref{GapEquationEnergy}).

To eliminate the singularity of the Fr\"ohlich potential, Bardeen et.al. approximated the Fr\"ohlich Hamiltonian by a box potential $V^{BCS}$ . Such a simplification is possible because the potential is integrated over in the gap equation, making the details of the potential insignificant. However, the price to be paid is the introduction of a fitting parameter $\omega_D$. This means, that the BCS potential can be used to explain superconductivity but not to predict it, because of the unknown $\omega_D$.

The last approach involves a more suitable renormalization procedure, which involves continuous unitary transformations. In this regard we mention the similarity renormalization first introduced by Glazek and Wilson~\cite{glazek1994perturbative} and the flow equations introduced by Wegner~\cite{wegner1994flow}. It has been shown \cite{mielke1997calculating} that the potential obtained through similarity renormalisation techniques $V^S$ can predict accurately the superconducting critical temperature.

\begin{figure}[h!]
  \centering
   \includegraphics[]{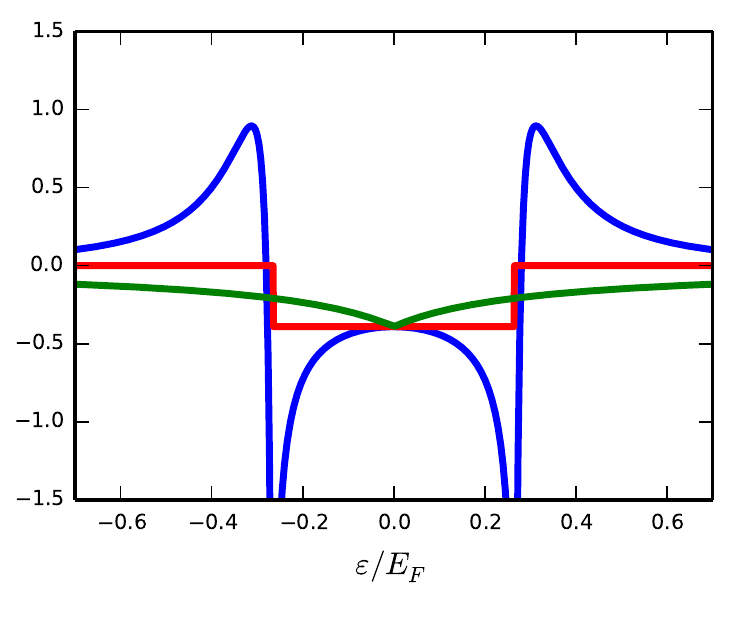}
  \caption{Comparison between $V^F(\varepsilon)$ (blue), $V^{BCS}(\varepsilon)$ (red) and $V^S(\varepsilon)$ (green). For the $BCS$ potential we choose $\omega_D=g_0$. For simulations we used typical GaAs parameters, the same parameters used for the solid lines in Figure~\ref{fig:RenormalizedPolariton}.}
   \label{fig:Comparison}
\end{figure}

To compare the different approaches we plot the three potentials $V(\varepsilon)$ on top of each other in Figure~\ref{fig:Comparison}. Notice that at the FS (i.e. $\epsilon=0$) all the potentials agree with each other, as they should since at this point the potential corresponds to real processes.

\subsection{Comparison to previous work}\label{sec:CompLaussy}
In the previous work on polariton-mediated superconductivity~\cite{laussy2010exciton,laussy2012superconductivity,cherotchenko2014superconductivity}, the authors used the Fr\"ohlich potential $V^F(\varepsilon)$. Notice that, although this potential is non-singular after being integrated, it still develops two large shoulders close to the Debye energy. The dependence of the critical temperature on the size and width of the shoulders has been investigated in Ref. \onlinecite{laussy2012superconductivity} and they have been used to predict a large critical temperatures obtainable in polariton-mediated superconductivity. As we discussed in the main text, we find much smaller $T_c$ for similar system parameters. Another consequence of the use of the Fr\"ohlich potential is the appearance of an oscillatory gap, which again has been treated as a peculiarity of polariton-mediated superconductivity. We argue on the other hand that the peculiarities mentioned above are by no means unique to polariton-mediated superconductivity. Instead, their appearance is due to the use of the Fr\"ohlich potential.

\bibliography{references}

\end{document}